\documentclass[preprint]{aastex}

\def\arcs{\char'175\ }
\def\arcsc{\char'175 }
\def\arcm{\char'023\ }
\def\arcmn{\char'023 }

\def\etal{et~al.\ }

\def\hub{\ifmmode H_\circ\else H$_\circ$\fi}
\def\kms{~km~s$^{-1}$\ }

\begin{document}


\title{RADIAL AGE AND METAL ABUNDANCE GRADIENTS IN THE STELLAR CONTENT OF M32}

\author{James A. Rose}
\affil{Department of Physics and Astronomy, University of North Carolina, Chapel
Hill, NC 27599}
\affil{Electronic mail: jim@physics.unc.edu}

\author{Nobuo Arimoto}
\affil{National Astronomical Observatory of Japan, Mitaka, Tokyo, Japan}
\affil{Electronic mail: arimoto@optik.mtk.nao.ac.jp}

\author{Nelson Caldwell}
\affil{F.L. Whipple Observatory, Smithsonian Institution, Box 97, Amado AZ
85645}
\affil{Electronic mail: caldwell@flwo99.sao.arizona.edu}

\author{Ricardo P. Schiavon}
\affil{Department of Astronomy, University of Virginia, 
P.O. Box 3818, Charlottesville, VA 22903-0818}
\affil{Electronic mail: ripisc@virginia.edu}

\author{Alexandre Vazdekis}
\affil{Instituto de Astrofisica de Canarias, La Laguna, Tenerife, Spain}
\affil{Electronic mail: vazdekis@ll.iac.es}

\author{Yoshihiko Yamada}
\affil{National Astronomical Observatory of Japan, Mitaka, Tokyo, Japan}
\affil{Electronic mail: yyamada@optik.mtk.nao.ac.jp}

\begin{abstract}

We present long-slit spectroscopy of the elliptical galaxy M32, obtained with 
the 8-m Subaru telescope at Mauna Kea, the 1.5-m
Tillinghast telescope at the F. L. Whipple Observatory,
and the 4-m Mayall telescope at the Kitt Peak National
Observatory.  The spectra cover the Lick index red spectral region
as well as higher order Balmer lines in the blue.  
Spectra have been taken with the slit
off-set from the nucleus to avoid scattered light contamination from the bright
nucleus of M32.   An analysis
of numerous absorption features, particularly involving the H$\gamma$ and
H$\beta$ Balmer lines, reveals that systematic radial trends are
evident in the integrated spectrum of M32.  Population synthesis models 
indicate a radial change in both the age and chemical composition of the
light-weighted mean stellar population in M32, from the nucleus out to
33\arcsc, i.e., approximately 1.0 effective radius, $R_e$.  Specifically, the 
light-weighted mean stellar population at 1 $R_e$
is older, by $\sim$3 Gyr, and more metal-poor, by $\sim$$-$0.25 dex in [Fe/H], than 
the central value of $\sim$4 Gyr and [Fe/H]$\sim$0.0.  We show that this
apparent population trend cannot be attributed to a varying contribution from
either hot stars or emission line contamination.  The increase in age 
and decrease in metal-abundance with radius are sufficiently well-matched
to explain the flat radial color profiles previously observed in M32.
In addition, the ratio
of Mg to Fe abundance, [Mg/Fe], increases from $\sim$$-$0.25 in the nucleus to
$\sim$$-$0.08 at 1 $R_e$.  Finally, we find spuriously pronounced line strength
gradients in the Mayall data that are an artifact of scattered
light from the bright nucleus.  Scattered light issues may explain the lack of
consistency among previously published studies of radial line strength
gradients in M32.

\end{abstract}

\keywords{galaxies: evolution --- galaxies: elliptical and lenticular, cD}

\section{Introduction}\label{Intro}

Due to its proximity and high central surface brightness, the elliptical galaxy
M32 has often served as a unique testing ground for evaluating the modelling of
the integrated light of galaxies and thereby disentangling key aspects of the
star formation and chemical enrichment histories of galaxies.  In fact, M32 is
still the only elliptical galaxy for which it is possible to obtain both
integrated light spectroscopy and an extensive CMD of the giant branch (using {\it HST}) at
the same location in the galaxy.  Thus it presents a powerful testing ground of 
the reality of integrated light analyses of galaxies.  Unfortunately, M32 
represents a rather special case elliptical, given its status as a low-mass
satellite galaxy to M31, its probable tidal truncation as a result of close
proximity to M31 (e.g., Bekki \etal 2001; Choi, Guhathakurta, \& Johnston 2002),
and its rare (cE) morphology.  Hence it is encouraging
that resolved star studies of the upper giant branch of
the more normal and massive elliptical galaxy NGC 3379 have been pursued
recently by Gregg \etal (2004).  Given its proximity, however, M32 remains the
ideal galaxy for testing inferences from integrated light spectroscopy and
broadband colors against resolved star observations.

In the context of constraining galaxy formation scenarios, determining radial
population gradients in elliptical galaxies can play a key role.  Specifically,
the predicted gradient in metallicity is substantially flatter in the case
of a hierarchical structure formation picture than it is for a monolithic
collapse scenario (e.g., Kobayashi \& Arimoto 1999).  Due to the high
surface brightness of M32, one can readily study the integrated spectrum
out to a radial distance of 30\arcsc, or 1 effective radius, R$_e$.  However,
previous studies of radial color and line strength gradients in M32 have led
to strikingly discordant results.  On the one hand, 
Peletier (1993) finds that all colors, as well as the CO index, remain constant
with radius, and concludes that no age gradient exists in M32.  Moreover,
in {\it HST}
images of the central regions of M32, Lauer \etal (1998),
find no gradient in V$-$I color, and only a very small blueing in U$-$V
with increasing radius.  In addition, {\it some} individual spectral features
exhibit no signs of a radial gradient in line strength.  Specifically,
both the Mg$_2$ index (Davidge 1991) and Mg~$b$ index (Gonz\'alez 1993; Fisher, Franx, \& Illingworth 1995)
appear to be constant with radius, as is Fe5015 (Fisher, Franx, \& Illingworth 1995).  
There is, however, considerable uncertainty concerning the flatness of the
Mg$_2$ radial profile, since Hardy \etal (1994) report a decline in this
index with radius out to a distance of 60\arcsc, or approximately 2 effective
radii, $R_e$.

On the other hand, 
Cohen (1979), Davidge, de Robertis, \& Yee (1990), and Davidge (1991) have 
found radial line strength gradients
in a number of spectral features, as have Fisher, Franx, \& Illingworth (1995) for the 
$\lambda$4668 feature.  The long-slit data of Gonz\'alez (1993), as reported in
Fisher, Franx, \& Illingworth (1995; cf., their Fig. 9), indicates a decrease in H$\beta$
strength with increasing radius. However, 
Hardy \etal (1994) find a flat profile in H$\beta$ out to 60\arcs in radius.
In addition, neither Fisher, Franx, \& Illingworth (1995) nor del Burgo \etal
(2001) find a radial trend in H$\beta$, although the latter two studies reach 
only to 20\arcs and 15\arcs respectively.  In addition, Ohl \etal (1998) have
found a strong far-UV (FUV) (1500\AA) color gradient in M32, such that the
FUV$-$B color becomes  bluer at increasing radial distance, which is the reverse
of the other early-type galaxies that they studied. Ohl \etal speculate that the
FUV gradient in M32 could be due to increasing age with radius, but also mention
the possibility of a varying helium abundance.  Such a UV color gradient, 
however, has {\it not} been found in recent GALEX observations (Gil de Paz \etal
2003).
As summarized by Davidge (1991), although there is considerable disagreement 
between investigators about the reality of line strength gradients in many (if not
most) features, a consistent interpretation of both line strength and color
behavior can be found if there is a radial gradient in M32 in both age and 
chemical composition, in the sense that the age increases with radial distance
while the metallicity decreases.

The most comprehensive study to date of the resolved stellar population of M32
has been carried out by Grillmair \etal (1996), who 
resolved individual stars down to slightly below the level of the horizontal
branch (HB) with {\it HST} WFPC2. The
observations cover an area at a radius of $\sim$1$-$2\arcmn ($\sim$2$-$4 $R_e$), with the PC chip field
centered at 58\arcs from the nucleus.  Grillmair \etal (1996) find a large 
spread in metallicity among the giant branch stars, although the
spread is less than that predicted in the simple closed box model of Searle \&
Sargent (1972).  The peak in the metallicity distribution appears to be near
solar, with a metal-weak tail extending to below [Fe/H]=$-$1.
A mean iron abundance of [Fe/H]=$-$0.25 is derived for an assumed
mean age of 8.5 Gyr, which is consistent with the results
inferred from modeling of the integrated
light indices of Gonz\'alez (1993), extrapolated out to the radius of the PC
chip.  In contrast, integrated light studies of the nucleus of M32 typically
indicate a mean age of $\sim$3$-$7 Gyr and [Fe/H]$\sim$0.0 (e.g., Jones \& 
Worthey 1995; Vazdekis \& Arimoto 1999; Trager \etal 2000; Worthey 2004;
Caldwell, Rose , \& Concannon 2003; Schiavon, Caldwell, \& Rose 2004).  
A number of studies have been made in the past of the AGB-tip stars in M32,
first from ground-based optical and IR imaging in good seeing (Freedman 1989, 
1992; Davidge \& Nieto 1992; Elston \& Silva 1992), and recently from adaptive
optics IR imaging from CFHT and Gemini-North by Davidge (2000) and Davidge \etal
(2000).  The earlier work found AGB-tip stars luminous enough to indicate the
presence of a substantial ($\sim$10$-$15\%) intermediate age population in M32,
while the adaptive optics studies have now pushed the same conclusion to a 
radial distance of only 2\arcsc.  As reported by Davidge \etal (2000), the 
implication of this recent work is that an intermediate age population is
uniformly spread over at least the central 2\arcm of M32, with no evidence
for a radial age gradient.  On the other hand, {\it HST} near-UV imaging of M32
by Brown \etal (1998) reveals that the resolved UV-bright stars are more 
centrally concentrated than the underlying unresolved UV sources, which 
indicates some type of radial population gradient.

It is clearly important to resolve the apparent discrepancy that exists between
color-based measurements and those based on individual absorption features
(as well as the implications of the resolved star imaging),
since many conclusions about the behavior of early-type galaxies are based on
colors alone (given that high S/N ratio long-slit spectroscopy is difficult to 
obtain for distant galaxies).  If radial population gradients in
age and metallicity exist in M32, in approximate proportions so as to cancel
out the primary color effects that each alone causes, as the line strength
measurements suggest, then the interpretation of color gradients in more distant
galaxies could be problematic.  With that issue in mind,
we present integrated light spectroscopy for M32
out to a radial distance of $\sim$30\arcs ($\sim$1 $R_e$). While not far enough to overlap with
the {\it HST} CMD study of Grillmair \etal (1996), a sufficient radial distance
from the nucleus is achieved to establish the existence of a significant radial 
population gradient
in M32, and thereby to link the CMD study with the numerous integrated 
light analyses of the nucleus.  
The emphasis of our study is on decoupling age and metallicity
effects using measurements of several Balmer lines.  

Here we combine long-slit spectroscopy from three telescope, spectrograph,
and detector combinations to evaluate the radial behavior of stellar populations
in M32.  The plan of the paper is as follows.  In \S 2 we describe the three
spectroscopic datasets, while the spectral indicators and populations
synthesis models employed are presented in \S 3.  In \S 4 we first establish
from two of the datasets that there is a small but significant radial change in
the light-weighted mean age and metallicity of M32 from the nucleus out to a
radial distance of 30\arcsc.  Then we demonstrate how scattered light from the
bright nucleus of M32 leads to an artificially strong apparent population 
gradient in the third dataset.  Some implications of the observed age/metallicity
gradient are briefly considered in \S5.

\section{Observational Data}\label{data}

\subsection{Subaru + FOCAS}\label{Subaru}

Long-slit spectra have been obtained of the elliptical galaxy M32 with the
8-m Subaru telescope at Mauna Kea with the FOCAS imaging spectrograph
(Kashikawa \etal 2000).  All spectra were obtained on the night of January 27,
2003.  The
spectra are imaged onto a 2K X 4K MIT CCD with 15 $\mu$ pixels, and binned
by a factor of two in the spatial direction, to achieve 0.2\arcsc/pixel.  A
slit width of 0.6\arcs was used, and the spectra cover the wavelength region
3960 $-$ 5560~\AA \ at a dispersion of 0.655 \AA/pixel, and a spectral
resolution of 93 \kms, or $\sim$3.1~\AA \ FWHM at 4200~\AA..  The slit was oriented
at a position angle of 80$^{\circ}$ E of N for all observations, which is only
10$^{\circ}$ from the minor axis, at $\sim$70$^{\circ}$ E of N (Tonry 
1984; Kent 1987 both report the PA of the major axis of M32 to be 
$\sim$$-$20$^{\circ}$ E of N).

The following exposures have been acquired.  First, two 300 second exposures were
made with the slit centered on the nucleus of M32.  Then, a series of three 900 second
exposures were made with the slit offset by 10\arcs from the nucleus, to moderate 
scattered light from the nucleus for analysis of the fainter outer region of M32. 
In addition, a 
900 second sky exposure was taken at a position offset by 6\arcm East from 
the nucleus of M32.  Note that the precise locations and PA of 
the slit for both the M32 and sky exposures is critical here, since we found
from both these, and earlier, observations that the slit can intersect one or
more planetary nebulae (PN) in M32 and/or a faint HII region in the outer disk
of M31.  We indeed find that our slit position centered on the nucleus of M32
intersects two PN that are $\sim$8 and 9.5\arcs along the slit to the NE,
specifically, PN \#24 and \#25 in Table 6 of Ciardullo \etal (1989).

Data reduction was carried out mainly with IRAF packages in the standard way for
long slit data; i.e. overscan correction, bias subtraction, flat fielding, 
$\lambda$ calibration, sky subtraction and flux calibration. 
Cosmic rays were removed and interpolated by using the ``cleanest'' task 
in the REDUCEME package (Cardiel et al. 1998). We initially ran automatic 
finding with the following parameters: SIGTHRESHOLD=6.0 in Searching Square=15. 
We took care to avoid mis-identification of CRs around the peak of flux in the
spatial direction. Then we checked by eye in three-dimensional surface plots of 
flux versus position on the CCD frame and edited out CRs found that way.
Since CRs can strongly affect the narrowly defined
H$\gamma_{\sigma<130}$ measurement, we checked whether 
CRs exist around  the H$\gamma$ feature. Fortunately we found no CR within
the passbands of the H$\gamma_\sigma$ index.

\subsection{Mt. Hopkins 1.5-m + FAST Spectrograph}\label{FAST}

Long-slit spectra of M32 have also been obtained with the FAST spectrograph
(Fabricant \etal 1998) on the 1.5m Tillinghast telescope of the Whipple Observatory, during the
night of 27 December 2002.  A 600 gpm
grating was used with a 3\arcs slit, which gave a spectral coverage of from 3600 to 
5500 \AA, at a resolution of 3.1 \AA \ FWHM, and a dispersion of 0.75 \AA/pixel on
the Loral 3K X 1K CCD.  The pixel scale along the slit is 0.57\arcs per
15$\micron$m pixel; however, the data were binned by 4 pixels along the slit,
to reduce the effective read noise in the outer low surface brightness regions
of M32. The detector has a read noise of 8.6 e$^-$ and a gain of
1.2 e$^-$/ADU.  A 300 second exposure was taken with the slit centered on the
nucleus of M32, at a PA of $-$20 E of N, i.e., with the slit aligned on the major
axis of the galaxy.  Then the telescope was moved sufficiently in Right 
Ascension (RA) so as to move the slit 5\arcs off the center of M32, and 
a 900 second exposure was taken.
Next, two 900 second exposures were taken with the
slit offset by 15\arcs from the center of M32, again with the repositioning
done in RA only.  These were sandwiched around a 900 second sky
exposure, with the slit offset by 10\arcm due South of the center of M32.  
Finally, a series of seven 900 second exposures were taken with the slit offset 
by 30\arcs from the center of M32.  Again the offset was made in RA, 
specifically, 2.9 seconds, which results in the slit being displaced by 30\arcs
from the center of M32.
Three more sky exposures and four arc lamp exposures were interspersed with 
these exposures, to provide reliable background subtraction and wavelength 
calibration.  Two of the sky exposures were offset by 15\arcm due South, while 
the third was offset by only 10\arcmn.  Given the major axis orientation of the
Tillinghast/FAST slit and the near-minor axis orientation of the Subaru slit, we have
sampled both principal axes of M32.

In additon to the above observations, we make use of a series of eight 4-minute
exposures on the nucleus of M32 taken on 28 December 2000 with the same instrument 
configuration as above.  The sole difference is that binning along the slit
direction was 3 pixels instead of 4 pixels per bin.  The total 32 minutes
integration provides superior S/N compared to the 5-minute nuclear exposure
on 27 December 2002, and the splitting of the integration into eight separate
exposures provides added information on errors in the spectral indices.

The Tillinghast/FAST spectra were processed as follows.  After bias-removal and flat-fielding, the
next step was cosmic ray (CR) removal.  For the 15\arcs and 30\arcs offset spectra,
the procedure is as follows.  A difference frame was created between two
individual exposures, a and b, with the difference frame given by a$-$b.  
In this difference frame, all pixels with counts below $-$100 were
flagged, and replaced with a 0, while all other pixels were set equal to 1.  Thus we
produce a mask frame with CRs flagged as zeros.  Then we ran a program that extends each
of the masked pixels by 1 pixel in each direction, to fully mask the wings of CRs.  Using
this revised mask file, the IRAF {\em fixpix} routine was run on frame b, to interpolate 
over the pixels contaminated by CRs.  This produced a CR-cleaned version of frame b.
To clean frame a, a b$-$a difference frame was created, and then the same procedure employed.
Using this technique, all frames were cleaned of CRs, except for the single short exposure
centered on the nucleus of M32, and the 5\arcs offset exposure.  The same procedure was 
followed to clean the sky frames of CRs.

The next step was to carry out sky subtraction.  We found that the arc lamp exposures did
not show zero point shifts greater than 0.3 pixels over the course of the M32 observations.
Consequently, we felt safe in directly subtracting the nearest sky exposure in UT to a
given M32 exposure, which simplifies matters.  We then extracted one-dimensional
spectra from the two-dimensional sky-subtracted, CR-cleaned data frames using
the IRAF {\em apall} task.  An aperture is defined, a trace is made of the
variation in the slit position of the aperture as a function of the
dispersion direction, and then the one-dimensional extraction is made based
on the trace.  Wavelength calibration was supplied by the nearest (in time) arc 
exposures.

\subsection{KPNO 4-m + R-C Spectrograph}\label{Mayall}

Long-slit spectra of M32 were acquired with the Kitt Peak National Observatory 
(KPNO) 4-m Mayall telescope by Dr. Lewis Jones, as part of his PhD 
thesis (cf., Jones 1999), on the nights of June 18-20, 1996.  The standard 
R-C spectrograph was used along with a 632 grooves/mm grating KPC-22B, in
second order, producing a dispersion of 0.7 \AA/pixel and a resolution of
1.8 \AA \ FWHM on the TK2B 2048$^2$ CCD detector.  The wavelength range covered
is 3300 \AA \ $-$ 4770 \AA.  The detector read noise and gain are 4 e$^-$ and
2 e$^-$/ADU respectively.  The full 5\arcm slit is covered at a scale of
0.69\arcsc/pixel; the slit width was 300 $\mu$m, or 2\arcsc.  Due to the late
rising of M32, and the desire to observe it at a minimum air mass, the
following strategy was used.  On each of the 
three nights of observation, two 1800 second
exposures were taken of M32 at the end of the night, preceded by an 1800 second 
sky exposure.  The sky exposure was taken at a Right Ascension that matched
the mean airmass of the
M32 observations.  However, on the first two nights the last M32 exposure 
extended far enough into the dawn twilight that the background level was 
substantially elevated; the M32 exposures were taken earlier on the third night,
and did not suffer from this problem.  We consequently have not used the last
M32 exposures on the first and second nights.
Thus a total of 2 useful hours of integration on M32 was obtained
through 4 individual exposures, and 1.5 hours of sky observation in three
exposures.  On each night the slit was rotated to the parallactic angle 
corresponding to the middle of the M32 observations.  The positon angle was
at 102\arcdeg, 100\arcdeg, and 155\arcdeg, respectively, for the first,
second, and third nights.  The position angle for the third night is,
coincidentally, within a few degrees of the major axis of M32.

Analysis of the above spectra was carried out in the NOAO IRAF software
package.  After initial trimming, bias-removal, and flat-fielding, 
the next step was to clean the numerous CRs from the individual 
sky exposures.  This was done in a two-step process.  First, the IRAF routine
'cosmicrays' was run, with the following parameters:
Threshold= 10.0, fluxratio=  5.00, window=7, npasses=5.  This approach removed
a majority of the CRs, but still left an unacceptable number in the
images.  Further lowering of the threshold for CR removal led to
removal of high points which are not CRs.  Instead, as a second step,
the 'median' routine was run, using a 1 x 10 median filter, with the large
binning factor oriented parallel to the slit.  This step is successful
at removing the remaining CRs, except for a few pathological cases which are
extended right along the slit direction.  The CR-cleaned sky exposures were 
then directly subtracted from the two M32 exposures from the particular night.

The next step involved removing the remaining CRs from the four individual 
sky-subtracted M32 exposures.  In addition, the data
frames needed to be transformed into a coordinate system in which the
slit axis and dispersion axis are orthogonal and oriented along columns and
rows.  The CRs were cleaned from the sky-subtracted M32 frames using the
same technique as described earlier for the FAST spectrograph data.  That is,
CRs were identified and eliminated from difference images between two
exposures.
A problem with this procedure is that the counts in the few
central rows of the M32 spectra are very high, and do not reproduce well from 
one exposure to another.  Thus many pixels in the central few rows are flagged.
To avoid losing all information in the cenral few rows, all of the flagged
pixels were subsequently reset to avoid interpolation.  As a result, there
is the possibility of badly positioned CRs to produce contamination in the
inner spectral indices.  We did, in fact, detect an instance of such a
contaminated spectral index, and recalculated it without the frame
which contains the CR hit.
Note that we could use the second M32 exposures on the first and second nights 
for the CR-cleaning purpose, since their backgrounds were only mildly elevated.

Following CR-removal, the four sky-subtracted M32 frames were
transformed into a coordinate system in which the slit axis and dispersion axis 
are orthogonal and oriented directly along columns and rows.
The transformation was carried out using the 'fitcoords' routine in
IRAF.  Then all four frames were averaged into a sky-subtracted 
wavelength-calibrated frame.  Note that this composite M32 long-slit
spectrum consists of two exposures at PA=101\arcdeg and two at PA=155\arcdeg.
Hence the final long-slit spectrum represents a crude azimuthal average of
the spectral behavior of M32.  From this final averaged M32 frame
spectra were extracted at various slit positions, and flux-calibrated using the
flux-calibrated spectrum of the nuclear region of M32 from L. Jones's (1999)
PhD thesis.  

\subsection{Extraction of Spectra}\label{extract}

One-dimensional spectra were extracted from the two-dimensional data frames 
using the IRAF `aptrace' routine to trace the deviation of the M32 spectra
from an idealized spectrum oriented strictly along the CCD rows.   In
performing the extraction (in the 'apall' routine) we used unweighted
extraction (as opposed to variance-weighted), since all CRs had been
identified and removed by this point.  To determine the light-weighted mean
radius of each extracted spectrum, we numerically integrated a radius-weighted
de Vaucouleurs r$^{1/4}$ surface brightness profile (de Vaucouleurs 1948)
over the extracted length
and width of the slit.  Following Kent (1987) we assumed a mean
ellipticity of 0.17 for M32, and effective radii along the major and minor axes
of 35\arcs and 29\arcs respectively.  Given that the Tillinghast/FAST data were taken with
the slit aligned parallel to the major axis of M32, the extractions out to a
light-weighted radius of $\sim$39\arcs correspond to $\sim$1.3 R$_e$ along the
minor axis. The Subaru slit was positioned very close to the minor axis, and
the data samples out to 30\arcsc, or 0.9 R$_e$, along the major axis.  A journal
of observations, specifying the PAs and locations of the various extractions,
is given in Table \ref{tab:journal}.

\subsection{Spectral Resolution}\label{resolution}

To make a consistent set of measurements of the spectral
indicators described in \S\ref{resolution}, it is necessary to carefully match the
spectral resolution at all positions along the M32 slit, and to determine
the radial velocity at each extracted slit position.  Resolution
variations along the slit can be due either to variation in the internal
velocity dispersion from the center outwards in the galaxy or to variations
in the spectrograph image quality. While velocity broadening produces an
effective resolution change that is contant in $\delta\lambda$/$\lambda$,
spectrograph image quality may vary in a complex manner both along
the slit and along the dispersion axis.  In what follows, we have
assumed that velocity broadening provides the dominant contribution to 
variations in spectral resolution along the slit.

The degree of velocity broadening for the M32 spectra extracted at various
slit positions was determined by cross-correlation with a
template spectrum, using the IRAF 'fxcor' task in the 'rv' package.  
The template is a model spectrum for a 3.55 Gyr solar abundance population from
Vazdekis (1999 and available at the URL 
http://www.iac.es/galeria/vazdekis/vazdekis\_models.html).  This model spectrum 
is created from a composite of
various stellar spectra in the Jones Coud\'e Feed Spectral Library, which is
available in the NOAO ftp archive (see below in \S\ref{otherdata}).  The 
resolution of the stellar spectra, and thus of our model template,
is 1.8 \AA \ FWHM.  This is different from the 3.1 \AA \ resolution typical
of our M32 observations.  The fxcor task
returns a measure of the FWHM of the cross-correlation peak.  To calibrate
this parameter, the template spectrum was Gaussian-smoothed by varying
$\sigma$'s and cross-correlated against the original unsmoothed spectrum.
The restricted wavelength region of $\lambda$$\lambda$4000 $-$ 4400 \AA \ was
used, thereby avoiding the Ca II H and K lines.  A fit
was made to the width parameter versus smoothing $\sigma$, in all cases by
making a 9-point Gaussian fit to the cross-correlation peak.  From a cubic
fit to peak width versus $\sigma$, a correspondence
between the two quantities was established.  Using
this calibration procedure, all of the extracted spectra of M32 at different
locations along the slit were smoothed to the same final spectral resolution.
Specifically, all extracted spectra were smoothed to a final Gaussian $\sigma$
of 130 \kms, which is in quadrature with the original 1.8 \AA \ FWHM of the
template spectrum.  Thus the final spectral resolution is equivalent to  
4.75 \AA \ at $\sim$4300 \AA.

\subsection{Other Observational Data}\label{otherdata}

In addition to the M32 long-slit spectra described above, we use two other
sources of data.  First, we use an integrated spectrum of the metal-rich 
Galactic globular cluster 47 Tuc obtained by A. Leonardi and J. Rose with
the CTIO 1.5-m telescope in November 1995.  An integrated spectrum of the
nucleus of M32 was also acquired with the same equipment on the same night, as
M32 transited the meridian, with the slit rotated to the parallactic N-S
position angle.  The characteristics of these spectra, which have a 
resolution of 3.1 \AA \ FWHM, are described in Leonardi \& Rose (2003). 
Hence a direct comparison is made between M32 and 47 Tuc with
the identical instrument setup, and can thus be tied in to the long-slit M32 
data.  

Second, we make use of two extensive libraries of stellar spectra
obtained with the coud\'e feed telescope at KPNO.  The first library, of 684 stellar
spectra, is publicly available at an NOAO FTP site 
(ftp://ftp.noao.edu/catalogs/coudelib/).  The characteristics of 
the spectra are decribed in Jones (1999) and in the readme file at the 
website.  In brief, the spectra were acquired at 1.8 \AA \ resolution (FWHM),
and cover two wavelength intervals, 3820 $-$ 4500 \AA \ and 4780 $-$ 5460 \AA.
We hereafter refer to this spectral database as the Jones Coud\'e Feed 
Spectral Library (JCFSL).  

The JCFSL spectra, as well as the M32 spectra and the 47 Tuc 
integrated spectrum, have been smoothed to the lowest spectral resolution of
3.1 \AA \ FWHM set by the CTIO 1.5-m 47 Tuc data.  In the case of the JCFSL,
for which resolution differences are due to spectrograph resolution rather than
velocity broadening, a gaussian smoothing by a $\sigma$ of 1.75 pixels in linear
wavelength is necesary to attain the resolution of the 47 Tuc integrated
spectrum.

The second spectral library, which became available while work for this paper
was in progress, is the more extensive Indo-US Library of Coud\'e Feed Stellar
Spectra (hereafter, IUCFSL), which includes 1273, covered at a spectral 
resolution of $\sim$1.2 \AA  \ FWHM. For most stars the entire region 3460 $-$
9464 \AA \ is covered.  The characteristics of this library, which is also
publicly available at an NOAO website (http://www.noao.edu/cflib/), are 
discussed in Valdes \etal (2004).  For these spectra, a correspondingly higher 
level of smoothing is required to attain the overall resolution of our M32 data.

Finally, to obtain a well-defined zero-point velocity determination, the radial
velocity for each of the extracted M32 spectra, and the 47 Tuc spectrum, was
determined using as template a population model spectrum based on a weighted composite of
JCFSL stars.  The method of producing the model spectrum, which has an age of 
4 Gyr and solar chemical composition, is discussed in \S\ref{models}.

\section{Spectral Indices and Population Models}\label{indmod}

\subsection{Spectral Indices}\label{sect:indices}

In this section we characterize the different types of spectral indices used
in our study.  Since these spectral indicators are all discussed extensively
elsewhere in the literature, we simply give brief summaries of their principal
characteristics and refer the reader to other sources for a more extensive
description.

There are a number of possible ways to isolate and measure an individual
spectral feature.  Here we rely on three different methods, which each have
particular advantages depending on the degree of line crowding, spectral 
resolution used, and importance of spectral resolution effects.  The first
type of measurement, which we refer to here as narrow-band equivalent width
indices, are the most familiar and have been widely used in the Lick system
of spectral indicators (e.g., Burstein \etal 1984; Worthey \etal 1994).  These
Lick equivalent width indices place a bandpass of typically 30$-$40 \AA \ 
(although some of the bluer indices have narrower bandpasses)
centered on a particular spectral feature, and then use bandpasses of similar
width on either side of the feature to provide a pseudo-continuum reference.
Due to their relatively broad width, the pseudo-continuum bandpasses often 
contain fairly strong absorption lines.  The
second system of measurements, which have been primarily developed in
Rose (1984; 1994), measure the relative depth of two neighboring features
without reference to the continuum.  This is accomplished by measuring the
residual central intensities in the two neighboring absorption lines, and 
forming their ratio.  Alternatively, a similar type of ratio can be measured
by using two neighboring pseudo-continuum peaks in the spectrum.  The term
``pseudo-continuum'' is employed here to emphasize that for integrated spectra
of galaxies, the true continuum is rarely measured; instead a local peak is
measured that comes closest to the true continuum value.  We refer to these
measurements as line ratio indices.  A third set of measurements involve 
isolating the pseudo-continuum peaks on either side of the absorption feature of
interest, and then directly measuring a pseudo-equivalent width.  These 
pseudo-continuum peaks, which are apparently free of absorption lines, at the
spectral resolutions typical of galaxy integrated spectroscopy, are defined
either by using the pixel with the highest counts as the local pseudo-continuum
peak or with an extremely narrow bandpass.  In this paper we make particular use
of this approach to isolate the H$\gamma$ absorption feature.  Variants of this
approach have been defined by Rose (1994), Jones \& Worthey (1995), 
Vazdekis \& Arimoto (1999), and Vazdekis \etal (2001b).  In the latter two 
references, great care has been taken to minimize the effect of spectral
resolution on the index, which is a major problem for this kind of index
(i.e., a pseudo-equivalent width is highly sensitive to slight changes in spectral
resolution).  In Vazdekis \etal (2001b), the metallicity sensitivity of the
index is reduced as well, by partially covering the neighboring metallic line
centered on $\lambda$ 4352 \AA.  This newest refinement on measuring the 
strength of H$\gamma$ is referred to as the  H$\gamma_{\sigma<130}$ index.

\subsubsection{Errors in Spectral Indices}\label{errors}

Errors in the observed spectral indices can arise from limitations in 
photon statistics, read noise, and flat field accuracy as well as from 
systematics of subtracting the sky background.  The optimal means of assessing
the random errors in spectral indices is from the scatter in values in repeat 
observations.  For the nucleus of M32 we have eight 4-minute exposures, and
thus can establish a robust determination of the standard error in the mean
index from the rms scatter.  Outside the nucleus, for each extraction at a
different radial distance we have divided the long-slit
observations into two non-overlapping regions on either side of the peak in the 
light profile.
We have then computed
the rms scatter in each spectral index from the pairs of observations on
opposite sides of the nucleus of M32 at the various radial positions.  Thus for 
each index we have an assessment of the random errors at each slit position
outside the nucleus from the scatter in only two independent measurements.  
Hence, the error for a particular index at a 
particular radial position has a considerable uncertainty from the above
approach.  Alternatively, we have computed the errors in spectral indices on the
basis of pure photon statistics and read noise in the CCD.  While this
approach is far more reliable statistically, it provides a less
comprehensive assessment of total uncertainties in the measurements.
Fortunately, we find that on average the photon statistical errors are 
generally consistent with the observed scatter from the paired observations.
As a result, the $\pm$1$\sigma$ errors on spectral indices cited 
in Tables~\ref{tab:Subaru} and \ref{tab:FAST} are
the more statistically secure errors derived from photon statistics.

We also calculated the potential effect of errors produced by uncertainties in
the background subtraction.  
All data reduction procedures as described above were repeated with the sky
exposures both enhanced and reduced by 10\%.  The spectral indices were
recalculated with the modified sky background subtraction, and the rms scatter
between the results obtained with enhanced and reduced sky values was 
calculated for each index.  The result is that even for the largest off-nuclear
radial position, 
the average error introduced by mis-subtraction of sky is significantly less 
than the errors from photon statistics.
Thus, even at the furthest radial position, and with a 
generous evaluation of the sky background uncertainty, the errors from uncertain
sky removal are considerably less than the random errors.  At smaller radial
distances, the sky errors will clearly be of even less significance.  The one
serious problem involved with background subtraction occurs if the slit for
the ``sky'' exposure or for the M32 exposure intersects a faint HII region in 
the disk of M31 an/or
a planetary nebula in M32 or M31.  The result is contamination of the Balmer
lines with emission.  We have in fact mentioned such problems, both on the
central Subaru slit (two PNe on the slit) and on a
Tillinghast/FAST background exposure.

Data on the spectral indices and their errors for the different radial slit
positions in M32 are listed in Table \ref{tab:Subaru} for the Subaru/FOCAS 
data and in Table \ref{tab:FAST} for the Tillinghast/FAST data.

\subsection{Population Synthesis Methods}\label{models}

A comparison of the observed index strengths to those predicted by 
stellar population synthesis models allow us to make a quantitative 
statement on the radial age/metallicity changes in M32. The most popular
approach followed by these models is to predict the strengths of
the strongest absorption features in galaxy spectra (Worthey 1994; 
Vazdekis et al. 1996). These models make use of polynomial fitting 
functions which relate the strengths of these features to stellar 
atmospheric parameters (i.e., T$_{eff}$, log~g, and [Fe/H]) on the 
basis of the Lick/IDS stellar spectral library (Gorgas et al. 1993; 
Worthey et al. 1994). 
A number of shortcomings of this approach prevent us from fully exploiting the 
information present in the high quality spectra of M32.
Among these problems is the fact that the spectral resolution of these models
(FWHM$>$8.5~\AA) is much lower than that intrinsic to M32. Moreover, the bluer
the wavelength the lower the resolution (e.g., FWHM$\sim$11.5~\AA\ for
$\lambda\sim$4000~\AA) (see Worthey \& Ottaviani 1997). Also, the fact that
the Lick/IDS stellar library was not flux-calibrated forces us to transform 
our data to the Lick/IDS instrumental response curve to be able to compare
the model predictions with the measured line-strengths. In addition, 
since these models only predict a specified number of spectral indices 
rather than the full spectra, we would be throwing away much of the 
information present in the M32 spectra, which is contained in the weaker 
spectral lines.

A further step in the modelling has been recently achieved by synthesizing 
full spectral energy distributions (SEDs) for old-age single burst stellar 
populations (SSPs) at a spectral resolution with FWHM=1.8\AA~(Vazdekis 1999;
Schiavon, Barbuy, \& Bruzual 2000;
Schiavon 2004) and flux-calibrated response. These newly developed models 
are based on the JCFSL, i.e., the extensive empirical stellar spectral library of 
Jones (1999) that we have described above. 
This new approach (see also Bruzual \& Charlot 2003) allows us to analyze a 
galaxy spectrum at a spectral
resolution limited only by its dynamics (i.e., its velocity dispersion, 
to which the model spectra should be smoothed). Thus, all the indices
described in \S\ref{sect:indices} can be measured directly on the model SEDs.
It is worth recalling that, thereafter, when we refer to the Lick indices 
we are not meaning those indices predicted by the previous models 
(with the specific instrumental response curve and resolutions of the 
Lick/IDS system), but we use the Lick index definitions for measuring 
these indices on the smoothed model SEDs (i.e., with flux-calibrated 
response and the resolution of M32).

In this paper we make use of two sets of models that differ in their 
essential ingredients. Whereas those of Vazdekis (1999) (as recently 
updated in Vazdekis \etal 2003) make use of the latest version of the 
Padua isochrones (Girardi et al. 2000), that of Schiavon (2004) uses 
the same isochrones and a new set of fitting functions to Lick/IDS
indices.  Furthermore, Vazdekis (1999) transforms the theoretical 
parameters of these isochrones to the observational plane following 
empirical stellar libraries such as those of Alonso, Arribas, \& 
Martinez-Roger (1996,1999) whereas Schiavon (2004) adopts the same
calibrations for F, G and K stars, and calibrations by Westera
et al. (2002) for cooler and hotter stars.  All of these models make 
use of scaled solar isochrones, thus do not directly model the non-solar 
abundance ratios found below.

\section{Radial Population Trends in M32}

\subsection{Evidence for an Age and Metal-Abundance Gradient}

\begin{figure}
\plotone{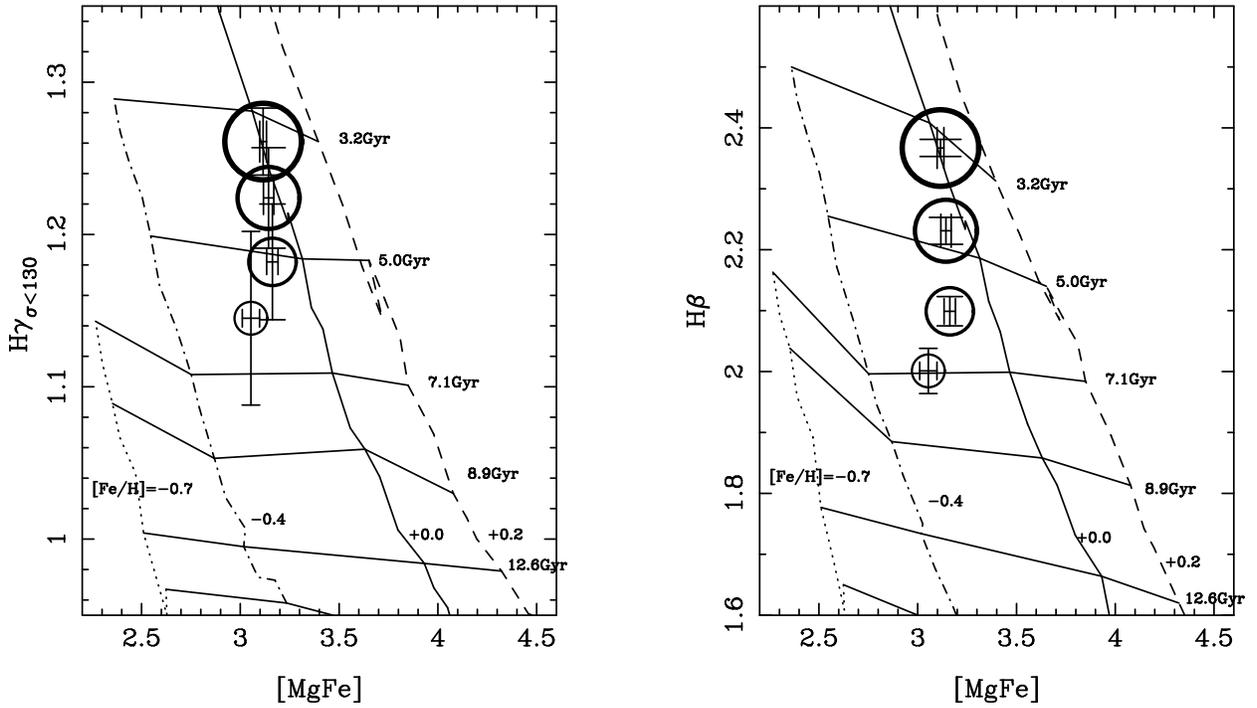}
\caption{The M32 H$\beta$ (right panel) and H$\gamma_{\sigma<130}$ (left panel)
indices are plotted versus the composite [MgFe] index for the Subaru data.  Overplotted on the
data points are the model grid lines of constant age and [Fe/H], with constant
age lines being nearly horizontal and constant [Fe/H] beng nearly vertical.  The
specific metallicities are, from left to right, [Fe/H]=-0.7, -0.4, 0.0, and 
+0.2.  The ages are, bottom to top, 17.8, 12.6, 8.9, 7.1, 5.0, and 3.6 Gyr.  The
data points, plotted as open circles with error bars are, from largest to
smallest, for the nucleus, 10\arcsc, 20\arcsc, and 30\arcs radial 
distance.  }
\label{fig:M32Sub}
\end{figure}

\begin{figure}
\plotone{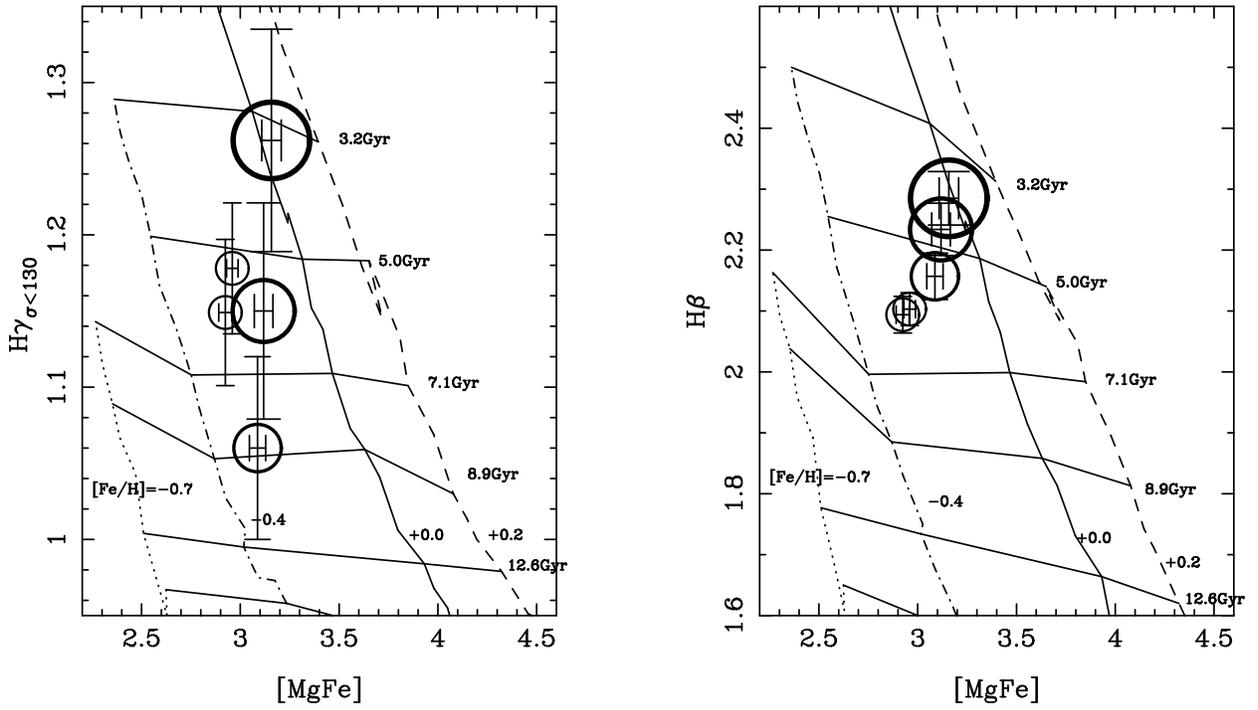}
\caption{The M32 H$\beta$ (right panel) and H$\gamma_{\sigma<130}$ (left panel)
indices are plotted versus [MgFe] for the FAST data.  
Model grid lines are the 
same as for Fig.~\ref{fig:M32Sub}.  
The data points, again having decreasing
size with larger radial distance, are at the nucleus, 7\arcsc, 19\arcsc,
35\arcsc, and 38\arcsc.
}
\label{fig:M32FAST}
\end{figure}

The central question of this paper is whether a population
gradient is present within the inner 30\arcs ($\sim$100 pc of M32.  We address this issue
in Figs.~\ref{fig:M32Sub} and \ref{fig:M32FAST}, where spectral index
data from the Subaru and Tillinghast/FAST observations, respectively, are plotted as a 
function of radial distance from the nucleus of M32.  To best disentangle
age from metallicity effects,
we have plotted spectral indices that measure the primarily age
sensitive Balmer lines versus an index that primarily responds to
mean metallicity.  Specifically, in Figs.~\ref{fig:M32Sub} and 
\ref{fig:M32FAST} the right hand panel is a plot of the Lick
H$\beta$ index versus the [MgFe] index, while the left panel is a plot of the
Vazdekis \etal (2001b) H$\gamma_{\sigma<130}$ index versus the same metallicity
sensitive [MgFe] index.  The [MgFe] index is a composite of both Fe-dominated
absorption lines and the 5175 \AA \ feature of the $\alpha$-element Mg, as
defined in Gonz\'alez (1993; ${\rm[MgFe]} = \sqrt{Mg~b\times (Fe5270 + Fe5335)/2}$), thus providing an assessment of the overall heavy
element abundance.  In fact, Thomas, Maraston, \& Bender (2003) find it to
be such a good tracer of overall metallicity that it is essentially independent
of [$\alpha$/Fe] (see also Bruzual \& Charlot 2003; Vazdekis \etal 2001a).   
In both plots we have included a grid of indices
based on the previously described single stellar population models of
Vazdekis (1999, and updated in Vazdekis \etal 2003) that cover
a range in age and [Fe/H].  The primarily horizontal lines connect models with
the same age while the primarily vertical lines connect models with the same
[Fe/H].  

The main result of Figs.~\ref{fig:M32Sub} and \ref{fig:M32FAST} is that a small age
and metallicity trend does indeed exist within the central $\sim$30\arcs (1 $R_{e}$)
of M32, in
the sense that the age increases from the center outward while the metallicity
slightly decreases.  Specifically, from the Subaru data plotted in 
Fig.~\ref{fig:M32Sub} it is evident that the light-weighted mean age 
increases from $\sim$3$-$4 Gyr in the nucleus to $\sim$6$-$7 Gyr at $\sim$30\arcs radius,
while the light-weighted mean metallicity (as measured by [MgFe]) 
decreases from almost exactly solar in the nucleus to $\sim$$-$0.3 at
30\arcsc.  There is good agreement between the results obtained from the 
H$\beta$ index and those from the H$\gamma_{\sigma<130}$ index.  The FAST
data, plotted in Fig.~\ref{fig:M32FAST}, provides similar results, particularly
with regard to the H$\beta$ versus [MgFe] plot.  The trend in 
H$\gamma_{\sigma<130}$ versus [MgFe] is less convincing, likely due to the
lower S/N ratio of the Tillinghast/FAST data.

\begin{figure}
\plotone{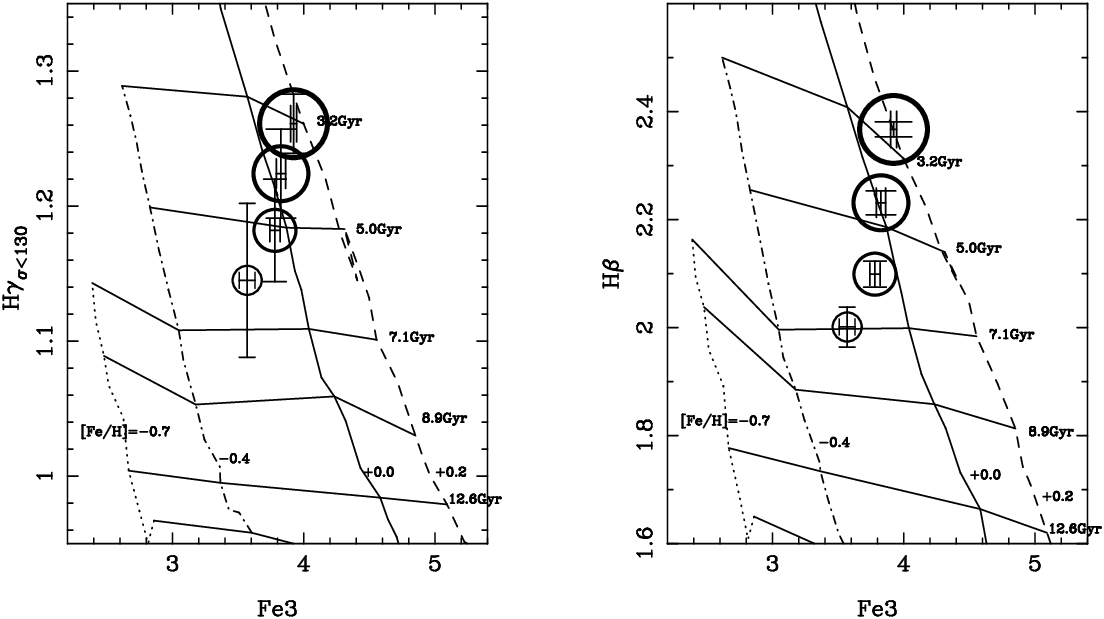}
\caption{The M32 H$\beta$ (right panel) and H$\gamma_{\sigma<130}$ (left panel)
indices are plotted versus the Fe3 composite index for the Subaru data.  Same
symbols and models grid lines as for Fig.~\ref{fig:M32Sub}.
}
\label{fig:M32Sub2}
\end{figure}

The overall trend in mean metallicity, as defined by [MgFe], can be further
refined to individual elements.  In Figs.~\ref{fig:M32Sub2}, \ref{fig:M32Sub3},
\ref{fig:M32Sub6}, and \ref{fig:M32Sub5} we plot the H$\beta$ and 
H$\gamma_{\sigma<130}$ indices versus the Lick Fe3 
(where Fe3 = (Fe4383+Fe5270+Fe5335)/3) , Mg~$b$, CN2, and G4300
indices, respectively, for the Subaru data.  Overall, the basic trend 
of decreasing
metal abundance with increasing radius is apparent.  However, as has been
discussed in Schiavon \etal (2004) and Worthey (2004), there are zero point
differences from one metal indicator to another with respect to the models
that indicate slight departures from a solar abundance ratio pattern in M32.
Specifically, the central values in M32 indicate that [Fe/H]$\sim$=+0.1 (as
seen in the Fe3 index in Fig.~\ref{fig:M32Sub2}), while [Mg/H]$\sim$=$-$0.1
(Fig.~\ref{fig:M32Sub3}), and the data for the CN2 index 
(Fig.~\ref{fig:M32Sub6}) indicates that [C+N/H]$\sim$+0.1.  In 
Fig.~\ref{fig:M32Sub5} the central value of the G-band, as characterized by the
G4300 index, implies that [C/H]$\sim$$-$0.2.  In addition, the size of the radial
gradient is different in some features, indicating that there is a radial
trend in abundance patterns as well.  We illustrate this fact in the case of
[Mg/Fe] by plotting the Mg~b index versus Fe3 in Fig.~\ref{fig:M32S6}.  
The M32 nucleus is
clearly shifted from the grid of solar abundance ratio models in the plot,
in the sense that Mg~b is too weak for the given Fe line strength.  However, at
larger radial distances there is a systematic shift of the M32 data towards the
model grid, indicating that [Mg/Fe] outside the nucleus shifts closer to the
solar abundance ratio.  We quantify the shift in [Mg/Fe] using the following
procedure.  Since the Mg~$b$ index is dominated by Mg and the Fe3 index by Fe
(Tripicco \& Bell 1995), we can extract approximate abundances for [Mg/H]
and [Fe/H] from the respective indices.  Specifically, plots of Mg~$b$ versus
both H$\beta$ and H$\gamma_{\sigma<130}$ provide nearly orthogonal age versus
metallicity model grids, allowing for a convenient approximation of [Mg/H]
(cf., Fig.~\ref{fig:M32Sub3}).  Similarly, plots of Fe3 versus both
H$\beta$ and H$\gamma_{\sigma<130}$ provide an approximation of [Fe/H]
(cf., Fig.~\ref{fig:M32Sub2}).  From the two individual abundances we
then infer [Mg/Fe].  Taken at face value, we find an
increase in [Mg/Fe] from $-$0.25 in the center of M32 to $-$0.08 at 30\arcs
radius.  Note that an alternative approach for obtaining the [Mg/Fe] ratios can 
be followed using models specifically computed for different 
$\alpha$-enhancements (e.g. Trager et al. 2000; Thomas et al. 2003), on the
basis of the sensitivities of these lines to the abundance changes of the
different species as tabulated in Tripicco \& Bell (1995).

The lower S/N ratio FAST data,
not plotted, shows essentially the same abundance patterns and trends as for
the Subaru data.  The slight
enhancement of Fe above solar, along with the underabundance of Mg 
and overabundance of CN with respect to solar, have also been reported in
Schiavon \etal (2004), based upon the same FAST data as used here, but 
incorporating Schiavon's (2004) population synthesis models.

To further illustrate the changing absorption line strengths with radius, in
Fig.~\ref{fig:ratiospec} we have plotted the ratio between the spectrum at
$\sim$30\arcs and that in the nucleus.  As can be seen in the ratio spectrum,
all major spectral features, including Ca II H and K and the Balmer lines
H$\delta$, H$\gamma$, and H$\beta$, appear in emission, indicating that these
features are deeper in the nuclear spectrum, as is expected if the central
population of M32 is both older and more metal-rich than at 30\arcsc.  That is,
the smaller (in equivalent width) Balmer lines at larger radial distance 
reflect an older age.

\subsection{Effect of Scattered Light on Observed Line Strength Gradients}

While the Subaru and Tillinghast/FAST data indicate only a very modest trend in age and 
metal-abundance in the central 30\arcs of M32, the KPNO 4-m data present a
very different story.  In Fig.~\ref{fig:M32_4m} the KPNO data are plotted
for a variety of spectral indicators versus the H$\gamma_{\sigma<130}$ age
indicator.  The Vazdekis model grids are again plotted, as in the previous
Figs.  Also plotted, as a representative population with age $\sim$12 Gyr and 
[Fe/H]$\sim$$-$0.7, is the integrated spectrum of the Galactic globular cluster
47 Tucanae.  The KPNO data indicate a strong radial age and metallicity 
gradient in M32, with the central age and metallicity in agreement with that
found in the Subaru and FAST data, but with the data at 30\arcs radius 
apparently having about the same age and metallicity as 47 Tuc.  Thus the
KPNO 4-m results are in striking disagreement with those found from the
Subaru and Tillinghast/FAST data, in that the gradient seen in the KPNO data is much more
pronounced than that found in the Subaru and FAST data.

A clue to understanding the large apparent line strength gradients in the
KPNO data is present in the behavior of the deepest lines in the spectrum, i.e.,
the Ca II H and K lines.  Since the integrated spectrum of M32 is overwhelmingly
dominated by cool stars, as is demonstrated below in \S\ref{hotstars}, the
depths of the Ca II H and K lines should not vary appreciably with distance
from the nucleus.  Instead, these lines have substantially lower depths in the
nucleus as opposed to in the outer regions, as is illustrated in 
Fig.~\ref{fig:depths}, where the spectrum at 30\arcs outside the nucleus is
normalized to and overplotted on the nuclear spectrum.  We quantify the
difference in Ca II H and K line depths seen in Fig.~\ref{fig:depths} by
using the pseudocontinuum peaks just shortward of Ca II K and longward of
Ca II H as references.  Specifically, the mean ratio between the line depth and
neighboring pseudocontinuum peak is 67\% smaller in the nucleus of M32 than it
is at 30\arcs outside the nucleus.

The elevated appearence of the Ca II line bottoms in the outer region spectrum
indciates that the Ca II lines (as well as all other spectral features) are
diluted by scattered light in the spectrograph.  Specifically, we speculate
that scattered light from the semi-stellar nucelus of M32, which reaches a
peak flux $\sim$100 times higher than at 30\arcs radius, contaminates the
spectrum at larger radii.  To further assess this conjecture we have generated a
featureless continuum with the same spectral response as for the raw observed
spectrum of the M32 nucleus.  We chose the scattered light component in this
manner because analysis of the scattered light spectrum of a star, HD187691, 
taken on the same night as the M32 data shows the spectrum of the scattered
light ot be a highly diluted version of the star's spectrum, with a raw flux 
distribution roughly similar to that of the star.  We added the above
featureless continuum in varying amounts to the M32
nuclear spectrum, up to a level that reproduces the line depths in Ca II H and
K at 30\arcs radius.  We have measured spectral indices in these artificially 
diluted spectra of M32, and find that the entire radial population gradient in
the KPNO 4-m data can be accounted for by scattered light contamination.  The
results of the scattered light simulation are plotted in Fig.~\ref{fig:scatter},
where the actual observed and simulated data are seen to overlap fully.  
Basically, dilution by scattered light weakens all spectral features, including 
both Balmer lines and metal lines.  Hence, one would expect scattered light to
produce an artificially older and more metal-poor spectrum, as is measured in
the KPNO 4-m data.

Given the evident problem of scattered light in the KPNO 4-m data, can we be 
sure that the more modest detected trend towards older and more metal-poor
mean population in M32 with increasing radius in the Tillinghast/FAST and Subaru data is
not also a spurious result?  We argue that the FAST and Subaru data are free
from scattered light problems becasue we have offset the slit from the nucleus
in the long exposure spectra which are used to extract information at larger
radii.  This is especially true for the FAST data, where we have obtained 
spectra with the slit offset by 5\arcsc, 15\arcsc, and 30\arcs from the
nucleus, so that our results always refer to spectra that are extracted from
regions where the flux is at the highest levels.  Clearly, the greatest 
scattered light problem arises when the nucleus, which is 100 times brighter
than at 30\arcs radius, is positioned on the slit.  Given the excellent
agreement between the Subaru and FAST data, we expect that by removing the 
nucleus from the slit in both datasets, we have basically alleviated the 
scattered light problem entirely.

To summarize, we have found a modest, but significant, radial population trend
in the central region of M32, such that the light-weighted mean age and
metallicity of M32 goes from $\sim$3$-$4 Gyr and [MgFe]=0.0 in the nucleus to
$\sim$6$-$7 Gyr and [MgFe]=$-$0.3 at 30\arcs radius.  A much stronger radial
population trend seen in the 4-m data is found to be an artifact of scattered
light in the spectra, produced by the location of the bright semi-stellar 
nucleus of M32 on the spectrograph slit.

\subsection{Comparison with the Results of Worthey (2004)}

In a recent preprint, Worthey (2004) also has obtained long-slit spectra for
M32, out to $\sim$45\arcs radius.  He finds the mean age and metallicity
for the nucleus of M32 to be 4 Gyr and +0.05, which is consistent with our
nuclear results.  He also finds that the mean age increases by a factor of
2.5 from the nucleus to 1 R$_e$, while the abundance drops by 0.3 dex.  Again,
our results are in good agreement with Worthey's, since we find the age
to increase by a factor 2, and abundance to decrease by 0.25 dex.  In addition,
Worthey (2004) finds that [Mg/Fe] is underabundant by $\sim$$-$0.1, which is
very close to our mean value of $\sim$$-$0.15.  However, he does not report a
radial gradient in [Mg/Fe] (we find that [Mg/Fe] inceases from $-$0.25 at the
nucleus to $-$0.08 at 1 R$_e$).  Finally, Worthey (2004) reports that CN indices
actually increase from the nucleus out to a maximum at 4\arcs radius before 
dropping at larger radii.  We instead see a monotonic decrease in CN from the
nucleus outward.  On the whole, the two studies are in excellent accord.

\subsection{Limits on the Role of Hot Stars}\label{hotstars}

The above results, taken at face value, indicate a $\sim$3 Gyr age gradient
within the central 30\arcs of M32.  However, it has been pointed out
previously (e.g., Maraston \& Thomas 2000; de Propris 2000; Brown \etal 2000) 
that a relatively
small population of hot stars (i.e., from a young hot star population, a 
metal-poor population, or a blue straggler population), if present in 
early-type galaxies, could ``contaminate'' the integrated spectrum, thereby
producing articially young light-weighted mean ages from models which
lack that component.  Thus we now consider whether the apparent age gradient
in M32, from $\sim$4 Gyr in the center to $\sim$7 Gyr at 30\arcs radius, could
instead result from the changing influence of a relatively small hot population.
To explain our case, there must exist a hot component in the center of M32 that
is less prevalent as one goes radially outward from the nucleus.

The hot star hypothesis can be tested by examining the behavior of the Ca II
index that was originally utilized by Rose (1985; 1994) and more extensively
modelled in Caldwell, Rose, \& Concannon (2003) and Leonardi \& Rose (2003).
The Ca II index, which is basically constant in
cool stars, declines very steeply towards early F and A stars as Ca II H and K
disappear and H$\epsilon$ (which is coincident in wavelength with Ca II H)
increases in strength towards the higher temperature
stars.  Specifically, the decline in the Ca II index sets in at T$_{eff}$ above
$\sim$8000 K for solar abundance stars.  As is summarized in Rose (1985; 1994) 
and Caldwell, Rose, \& Concannon 
(2003), a small deficit in the Ca II index, compared with the ``expected''
value of 1.20 for a strictly old population, is observed in the nucleus
of M32.  The deficit of $\sim$0.05 from the ``expected''
value indicates that $\sim$3\% of the light at
4000 \AA \ in M32 is provided by hot (i.e., A-type) stars.  This population
fails by a factor of 5 to account for the overall strong Balmer lines
observed in M32, i.e., an intermediate-age ($\sim$4 Gyr old) population is
required to explain the integrated spectrum of M32.  Here we consider whether
a gradient in the hot population might provide an explanation for the
$\sim$3 Gyr age gradient observed in the central 30\arcs of M32.

Unfortunately, the Subaru spectra do not extend blueward enough to provide
coverage of Ca II K, and the KPNO 4-m  spectra are compromised by scattered 
light, hence we rely exclusively on the FAST data.   The Ca II index at various
radii in M32 is plotted in Fig~\ref{fig:caII}.  Within the observational errors
the Ca II index is seen to be roughly constant, indicating that the $\sim$0.05
deficit in the index due to a hot star population is present in essentially
constant proportion from the nucleus through a radius of 30\arcsc.  To 
realize the implication of the small, constant Ca II deficit in M32, we have
done the following exercise.  We have both added and subtracted in varying
amounts the spectrum of the A4V star HD136729 to the central Tillinghast/FAST spectrum of
M32.  Here we have used the spectrum of HD136729 available in the IUCFSL,
and have rebinned it and smoothed it to match that of the M32 data.  We have
also normalized the HD136729 to that of M32 at 4000 \AA.  Specifically, we
added and subtracted the spectrum of HD136729 at 5\%, 10\%, and 20\% of the
level of M32 (at 4000 \AA).  The effect on the Ca II index of M32 can be seen
in Fig~\ref{fig:caII}.  Subtracting or adding only a 5\% contribution from
HD136729 at 4000 \AA \ changes the Ca II index of M32 by $\pm$0.09, which is
substantially higher than the observed deficit of $\sim$0.05 of M32 from the
``standard'' value of 1.2 for an entirely cool star population.  Thus only
$\sim$3\% of the light at 4000 \AA \ in M32 comes from hot stars.  Furthermore,
this contribution is constant to within $\sim$$\pm$0.02 in the Ca II index over
the central 30\arcs of M32.  If we now consider the impact of a 3\% contribution
of A4V light at 4000 \AA \ on the H$\beta$ index, we find that the H$\beta$ index
changes by only $\pm$0.09 as the 3\% A4 star contribution is added and 
subtracted.  As can be seen in Fig.~\ref{fig:M32Sub}, a change in H$\beta$ of
0.09 produces only a tiny effect on the derived age.  Given that we find the
Ca II index to vary radially by less then 0.02, we thus infer that a gradient
in H$\beta$ due to a gradient in hot star contribution can be present at no more
than the 0.04 level in that index, which is far below the observed H$\beta$
radial gradient of $\sim$0.4.

Finally, we have repeated the above analysis using the solar abundance 0.32
Gyr model population from Vazdekis (1999), and obtain very similar results as
were obtained with the A4V star, since the 0.32 Gyr model spectrum is 
similar to that of the A4V star HD136729 in the blue.  Thus we also conclude 
that at most only $\sim$3\% of the light at 4000 \AA \ in M32 comes from a 
$\sim$0.3 Gyr population.

\subsection{Limits on Contamination from Emission Lines}\label{emission}

Another important issue to consider is the possibility that contamination from
emission lines could be systematically affecting the results derived from the
absorption line indices.  The most obvious concern is emission fill-in for
the Balmer lines.  Here we are primarily interested in the potential effect
of a radial {\it gradient} in emission producing spurious population effects.
Since the data indicate that 
age systematically increases with increasing radial distance from the
center of M32, we must assume then that emission contamination also increases
with radial distance, if it is producing a spurious age increase due to emission
fill-in of the Balmer lines.  Such a scenario is perhaps somewhat contrived, in
that one might expect such emission to be centrally concentrated, but we will
consider that possibility nonetheless.

There are three potential sources of emission contamination.  The first occurs
if the spectrograph slit intersects a planetary nebula either, in M32 or M31, 
and can contaminate either the slit position through M32 or the offset ``sky''
slit position.  The second possibility is that the slit can be positioned on
an HII region in M31.  Again, this could be either the slit positioned on M32
or at the offset sky position.  Finally, there can in principle be a weak AGN
emission spectrum in the nucleus of M32.

All three sources of emission are plausible problems.  Ciardullo \etal (1989)
report positions for seven planetary nebulae (PNe) within the central 30\arcs of
M32.  As mentioned earlier, the Subaru spectrum through the nucleus of M32, at
a PA of 80\arcdeg E of N in fact intersect two of these PNe's, \#24 and 25
in the Ciardullo \etal (1989) Table 6.  Inclusion of these two PNe lead to a
very significant fill-in of the Balmer lines at 10\arcs radius.  We have
also found that a sky position for spectra taken of M32 with the FLWO 1.5-m
and the FAST spectrograph that are not used in this paper is contaminated with
a faint HII region from the M31 disk on part of the slit.  In that case we 
observed an unexpected enhancement in the H$\beta$ absoprtion strength in M32,
since H$\beta$ absorption in our sky position was artificially lowered due to 
fill-in from the HII region emission.  Finally, M32 is known to harbor a
central mass concentration (Tonry 1984; Dressler \& Richstone 1988; van der 
Marel \etal 1994, 1998; Verolme \etal 2002).  Thus an AGN spectrum is not
unexpected.

The possibility that emission fill-in is an important effect in the nucleus of
M32 (and for other low-luminosity early-type galaxies) has been discussed 
in Jones (1999).  Here we briefly summarize the conclusions.  Gonz\'alez (1993)
detected a slight amount (0.11 \AA \ in equivalent width) of [OIII]$\lambda$5007
emission in the nuclear spectrum of M32.  Jones's (1999) spectra do not cover
[OIII]$\lambda$5007, but provide tight upper limits on 
[OII]$\lambda$3727, from which strong constraints can be set on the type of
emission spectrum that is contributing to M32.  In particular, a LINER spectrum
is decisively ruled out, since, based on the observed [OIII]$\lambda$5007
emission of 0.11 \AA \ in equivalent width found by Gonz\'alez (1993), one 
predicts an equivalent width of [OII]$\lambda$3727 in excess of the observed
upper limit.
Instead, the two possibilities that remain are either 
planetary nebula emission or emission from an HII region.  In both cases, the
amount of emission predicted at H$\gamma$ from the small [OIII]$\lambda$5007
detection by Gonz\'alez (1993), is only 0.008 \AA \ in the planetary nebula
case and 0.005 \AA \ in the HII region case, which leads to an age
adjustment of $<$1 Gyr.  Note that the adjustment is such that emission tends
to make the galaxy look {\it older} than it actually is.

We have reinvestigated the nuclear spectrum of M32 using the eight 4-minute
exposures acquired with the FAST spectrograph.  These spectra cover both
[OIII]$\lambda$5007 and [OII]$\lambda$3727.  A key issue in emission line
detection is to find an appropriate template spectrum, which matches the
absorption features of M32 but is certified to be without emission 
contamination itself.  For a template we have used the synthesized integrated
spectrum of the Galactic cluster M67 from Schiavon \etal (2004).  In the
Schiavon \etal (2004) paper M67 was found to have a very similar age and
chemical composition to the luminosity-weighted mean age and metallicity of
M32.  Since the M67 integrated spectrum is built up from coadding spectra of
individual stars in M67, the resultant integrated spectrum is certainly free
from emission.  In addition, the M67 spectra were acquired with the same FAST
spectrograph and grating combination as for M32.  We have registered
the M67 spectrum to that of M32 and formed the ratio spectrum between the two.
The ratio spectrum is plotted in the region of [OII]$\lambda$5007 and
[OIII]$\lambda$3727 in Figs.~\ref{fig:emis5007} and \ref{fig:emis3727}
respectively.
We do not see any conclusive signs of either [OIII]$\lambda$5007 or 
[OII]$\lambda$3727.  Rather, we establish an upper limit to the
equivalent width of [OIII]$\lambda$5007 of 0.05 \AA, and a limit to 
[OII]$\lambda$3727 of 0.15 \AA.  The fact that our upper limit on 
[OIII]$\lambda$5007 conflicts with the detection by Gonz\'alez (1993) can be due
to the different apertures used for the ``nucleus'' of M32.  In any case,
emission contamination of the Balmer lines in the center of M32 is small.

Given that we have established the nuclear spectrum of M32 to contain little
or no emission contamination, we now use the nuclear spectrum of M32 as a
template and divide it into the other spectra extracted at different radii.
We then look for [OII]$\lambda$3727 and [OIII]$\lambda$5007 emission in these 
template-normalized spectra.  Due to the radial gradient in the absorption
spectrum of M32, 
using the nuclear spectrum of M32 as a template is not completely
successful at cleanly removing the underlying absorption spectrum in the
neighborhood of [OII]$\lambda$3727 and [OIII]$\lambda$5007 at extra-nuclear 
positions in M32.  This problem is particularly apparent at the blue end of
the spectrum, where all of the key spectral features (such as Ca II H and K)
appear in emission, since they are deeper in the M32 nucleus, as well as in
the Balmer lines.  However,
applying a conservative upper limit defined from the spurious peaks in the
neighborhood of [OII]$\lambda$3727 we conclude that the equivalent width of
[OII]$\lambda$3727 emission is below 0.2 \AA \ FWHM at all locations along
the slit out to $\sim$30\arcsc, and the limit on [OIII]$\lambda$5007 is
$\sim$0.05 \AA.  For high-excitation emission spectra, such as from PNe and
metal-poor HII regions, the limit on [OIII]$\lambda$5007 emission produces
a stringent limit on H$\beta$ emission.  On the other hand, for low-excitation
HII regions, the [OII]$\lambda$3727 emission
exceeds H$\beta$ emission, thus the limit on [OII]$\lambda$3727 emission
produces a strict limit on H$\beta$ emission.  In short, we find that emission
line fill-in in the outer region of M32 cannot account for the observed age
and metallicity trend with radius.  As well, if the gradient were due to
discrete PNe and/or HII regions on the slit, we would readily detect them on
the original two-dimensional spectra, as we have done in a couple of cases.

\section{Discussion}\label{discuss}

The results from the previous section suggest that there is a radial gradient
in the mean age and metallicity in the central regions of M32, such that the
mean age rises from $\sim$3$-$4 Gyr in the nucleus to $\sim$6$-$7 Gyr at 30\arcs
radius, while the metallicity drops from [Fe/H]$\sim$0.0 to $\sim$$-$0.3.  
In addition, we infer a slight gradient in [Mg/Fe] such that [Mg/Fe] increases
from $-$0.25 in the nucleus to $-$0.08 at 30\arcs radius.  These age and
chemical enrichment trends all point towards a more extended history of star 
formation and chemical enrichment in the nucleus of M32 than at 1 R$_e$.  
Of particular note is that the higher mean overall metallicity in the center 
of M32, coupled with the lower ratio of the 
$\alpha$-element Mg relative to Fe, imply both a greater chemical
enrichment in the nucleus as well as a longer time period for that
enrichment, since [$\alpha$/Fe] may be a reflection of relative timescales of
Type II versus Type Ia supernovae.  In short, a consistent picture of the
radial line strength trends in M32 emerges from this study.

\begin{figure}
\plotone{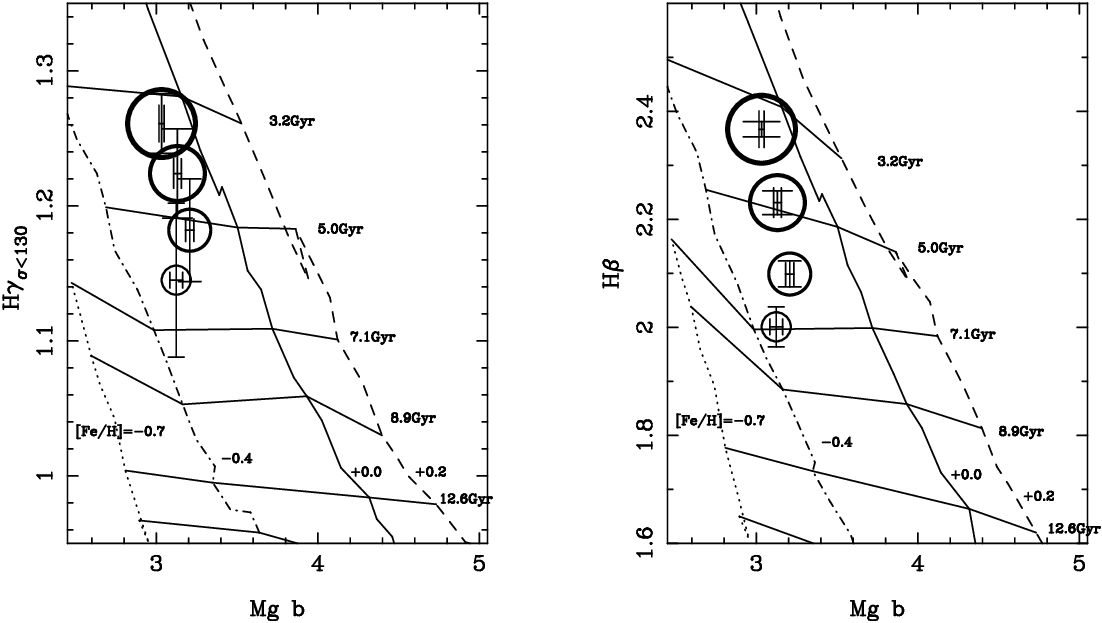}
\caption{The M32 H$\beta$ (right panel) and H$\gamma_{\sigma<130}$ (left panel)
indices are plotted versus the Mg~$b$ index for the Subaru data.  Same
symbols and models grid lines as for Fig.~\ref{fig:M32Sub}.
}
\label{fig:M32Sub3}
\end{figure}

\begin{figure}
\plotone{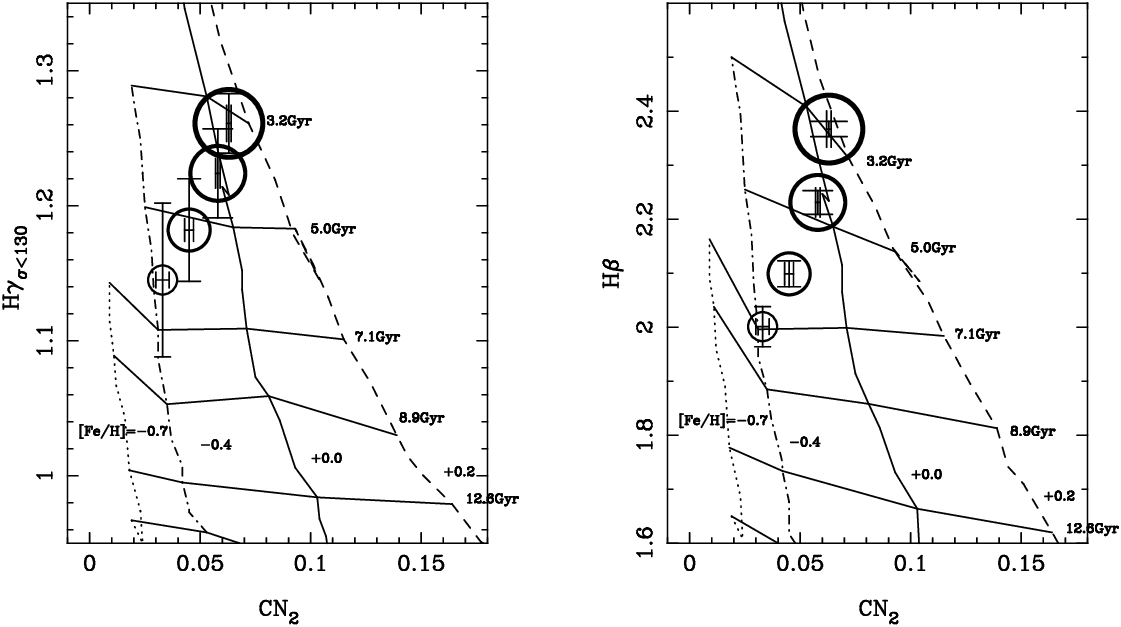}
\caption{The M32 H$\beta$ (right panel) and H$\gamma_{\sigma<130}$ (left panel)
indices are plotted versus the CN$_2$ index for the Subaru data.  Same
symbols and models grid lines as for Fig.~\ref{fig:M32Sub}.
}
\label{fig:M32Sub6}
\end{figure}

\begin{figure}
\plotone{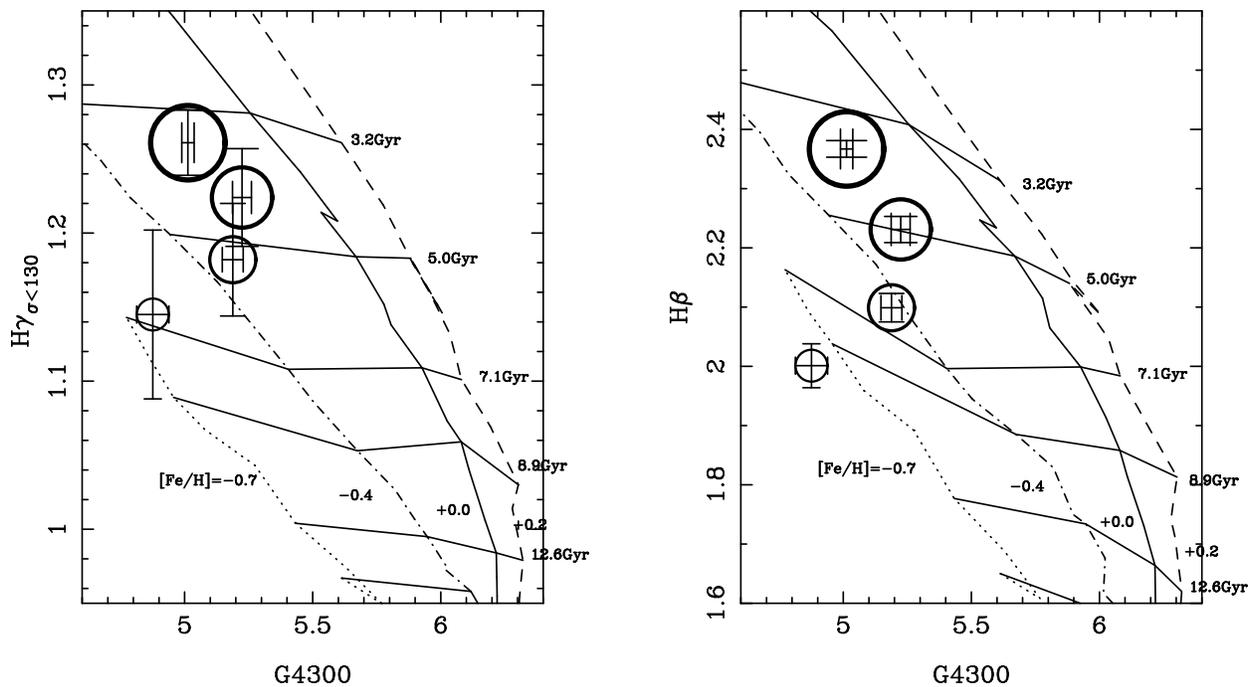}
\caption{The M32 H$\beta$ (right panel) and H$\gamma_{\sigma<130}$ (left panel)
indices are plotted versus the G-band index for the Subaru data.  Same
symbols and models grid lines as for Fig.~\ref{fig:M32Sub}.
}
\label{fig:M32Sub5}
\end{figure}

\begin{figure}
\plotone{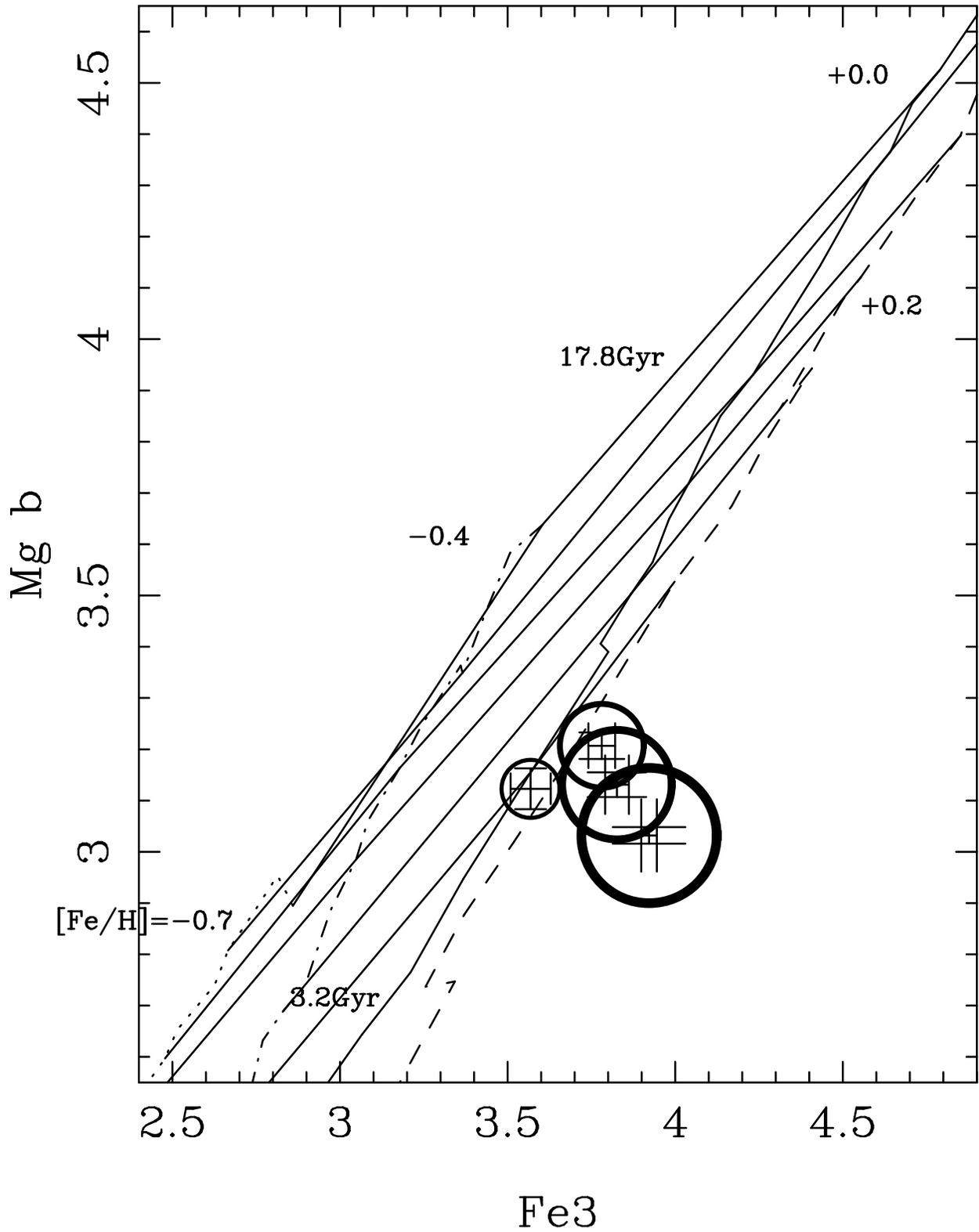}
\caption{The M32 Mg~$b$ index is plotted versus the Fe3 index for the Subaru
data. Same symbols and models grid lines as for Fig.~\ref{fig:M32Sub}.
}
\label{fig:M32S6}
\end{figure}

\begin{figure}
\plotone{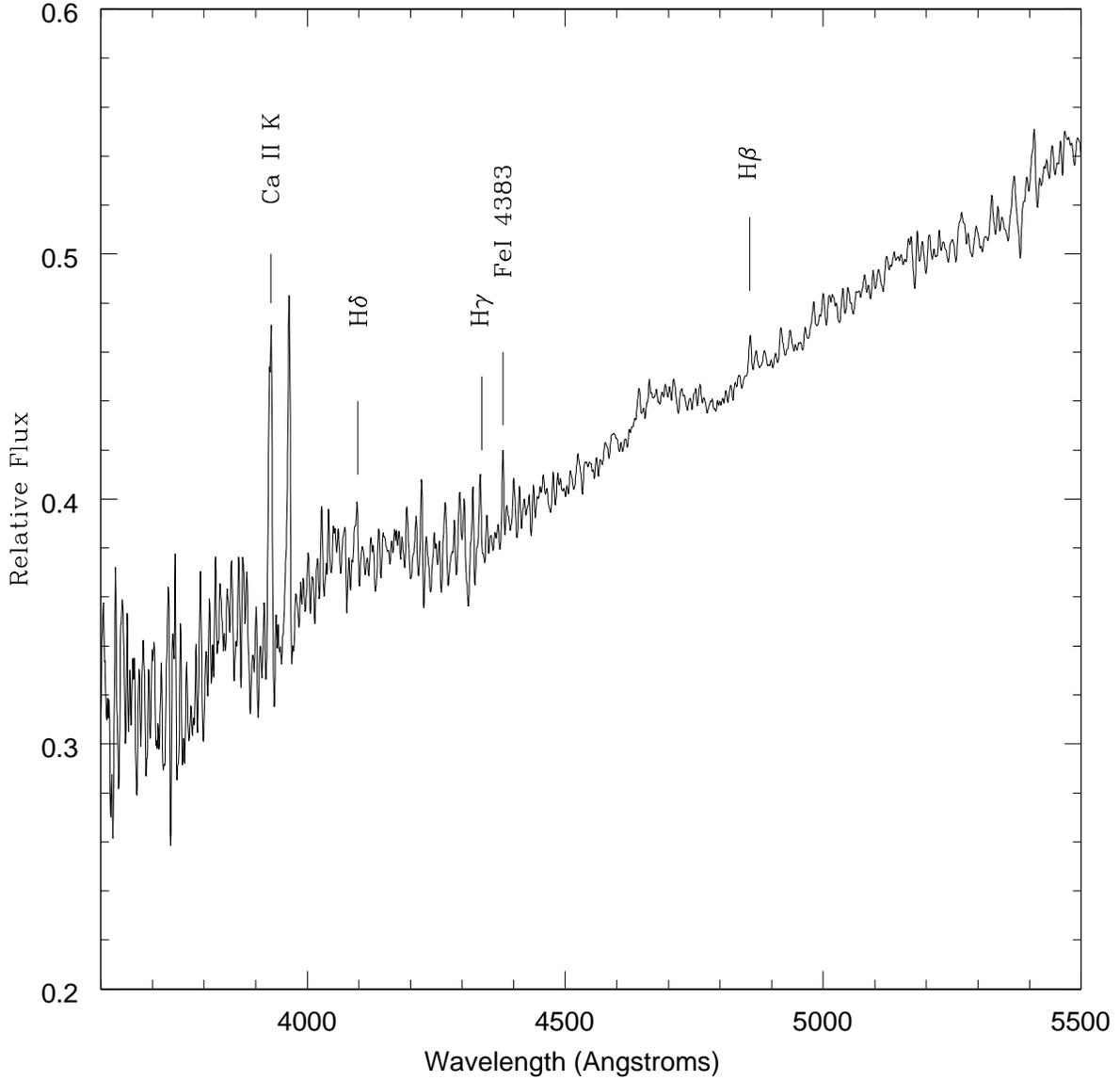}
\caption{The spectrum at 30\arcs radius is divided by that in the nucelus for 
the FAST data.  This ratio spectrum demonstrates that all principal spectral
features, some of which are specifically marked, show up in emission, thus
indicating that they are stronger in the nucleus of M32.  Note that the large
overall color gradient is probably an artifact of extracting the nuclear 
spectrum with a small aperture.}
\label{fig:ratiospec}
\end{figure}

\begin{figure}
\plotone{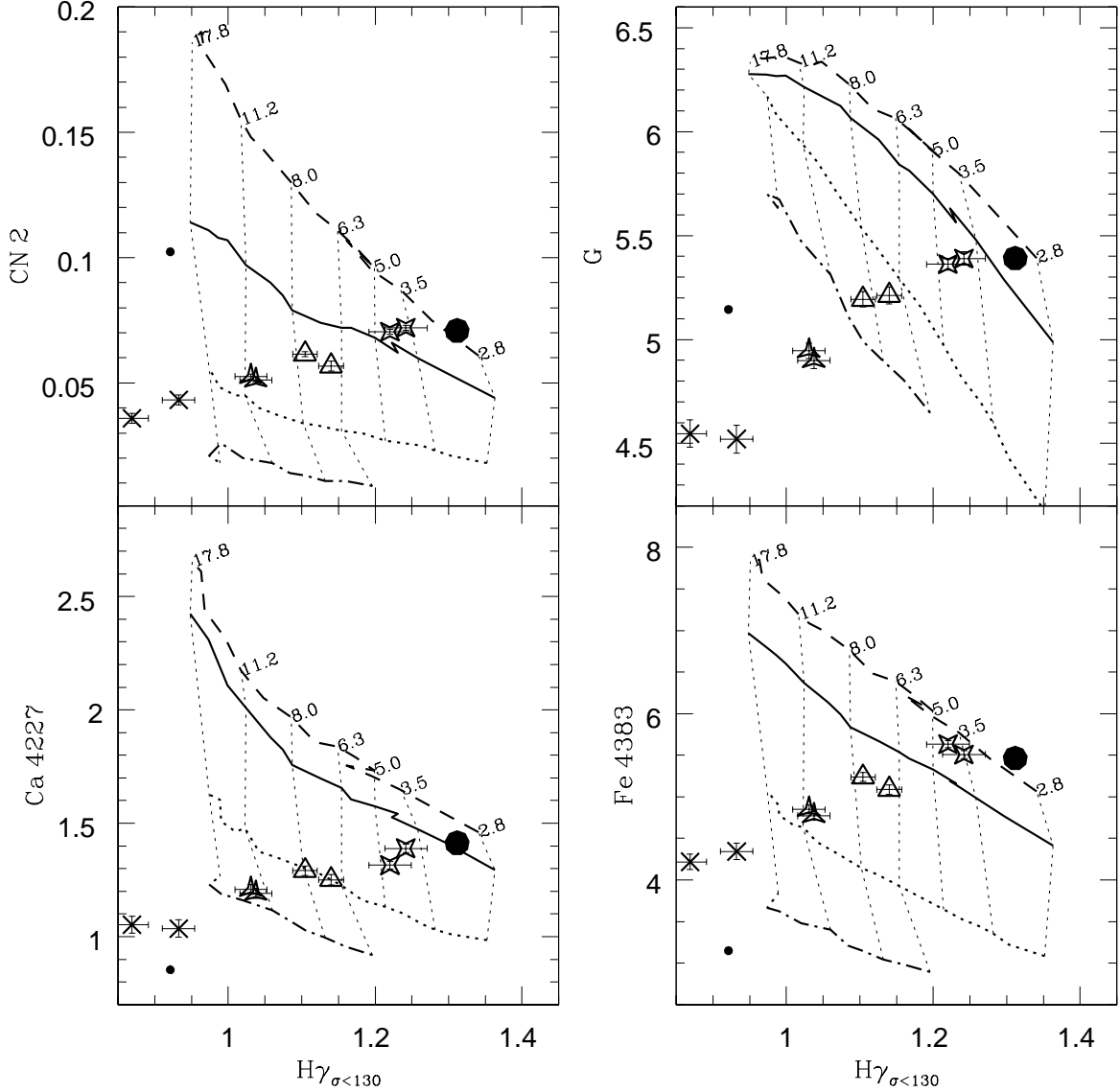}
\caption{Various Lick spectral indices are plotted versus the 
H$\gamma_{\sigma<130}$ index for the Mayall telescope data for both M32 and
for the Galactic globular cluster 47 Tuc.  The small filled circle denotes
47 Tuc.  The large filled circle represents data for the M32 nucleus. The
pairs of starred squares, open triangles, starred triangles, and x's represent
data at radial distances of 4.1\arcsc, 8.6\arcsc, 19.3\arcsc and 34.5\arcsc, respectively.
Also plotted are model grid lines, with ages marked on the nearly vertical
grid lines, ranging from 2.8 Gyr to 17.8 Gyr.  Horizontal lines of constant
metallicity are, from top to bottom, [Fe/H]=+0.2, 0.0, -0.4, and -0.7.
}
\label{fig:M32_4m}
\end{figure}

\begin{figure}
\plotone{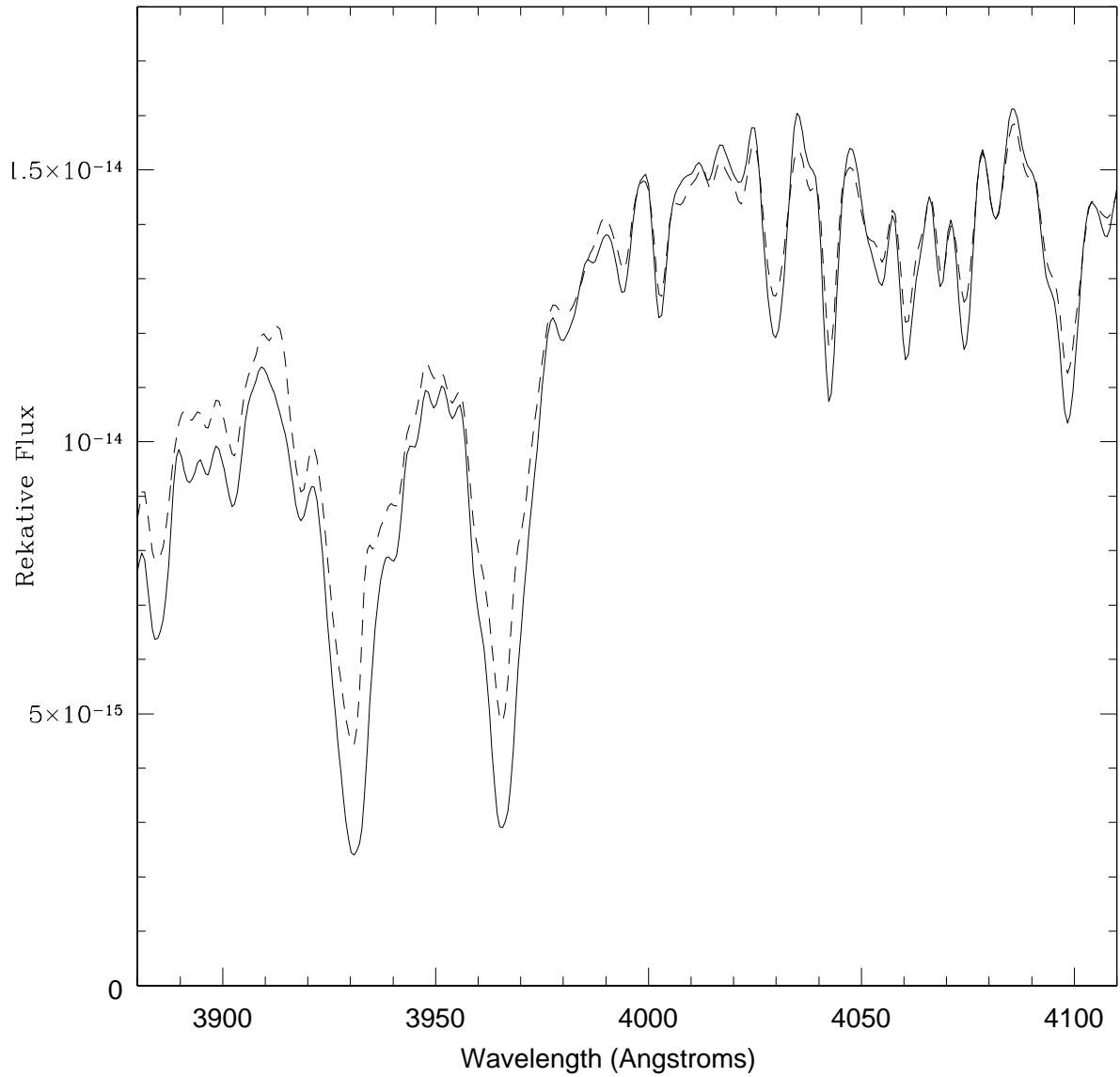}
\caption{The nuclear spectrum of M32 (solid line) acquired with the Mayall
telescope is overplotted with the
spectrum at 30\arcs (dashed line).  The spectrum at 30\arcs has been 
multiplied by a factor of 100.  Note the shallower Ca II H and K lines in the
spectrum at 30\arcsc, due to scattered light from the nucleus.}
\label{fig:depths}
\end{figure}

\begin{figure}
\plotone{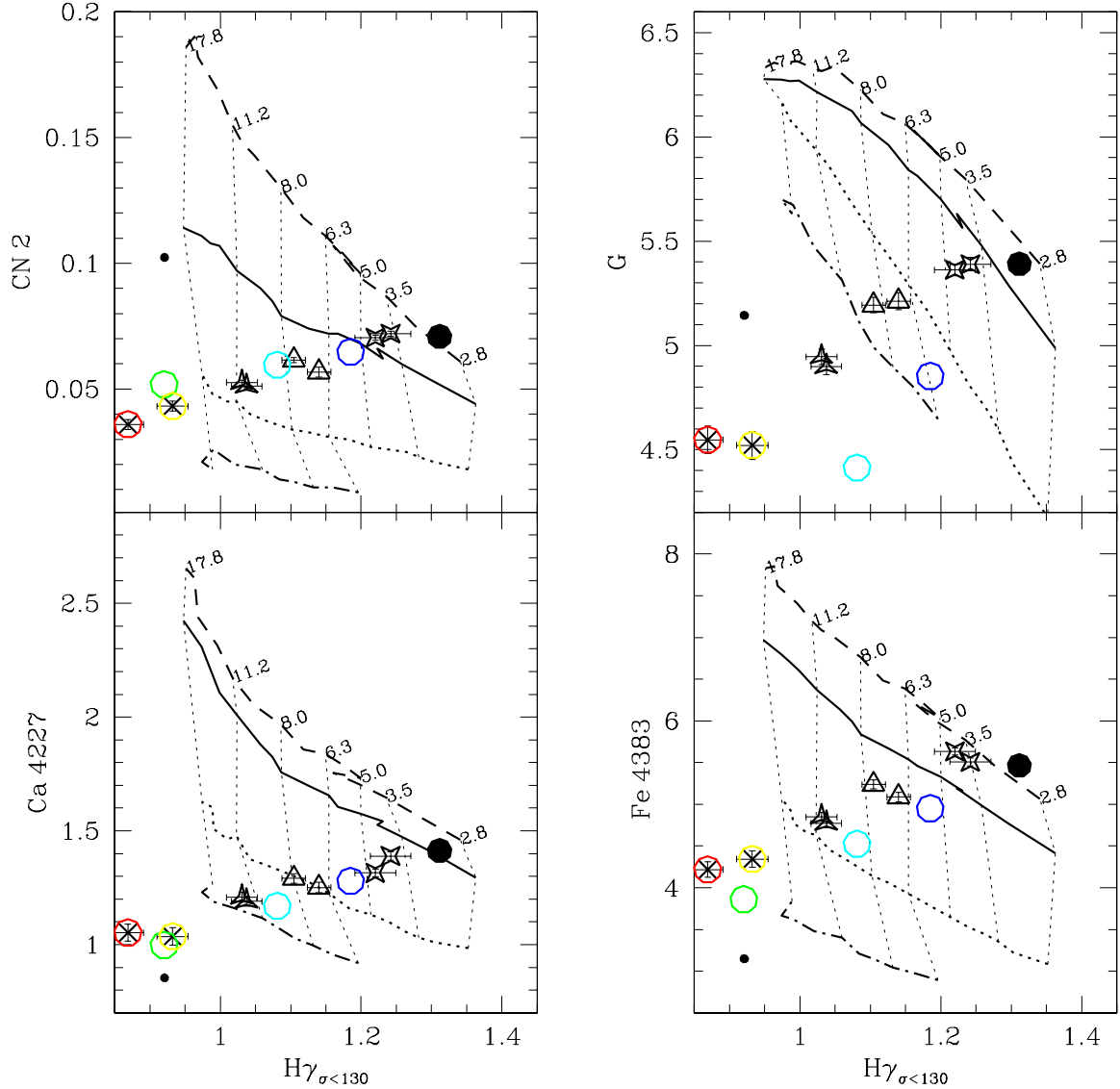}
\caption{The same spectral index diagrams as in Fig.~\ref{fig:M32_4m} are
are replotted.  As in Fig.~\ref{fig:M32_4m}, the symbols with error bars
are the M32 data while 47 Tuc is represented by a small filled circle.  All
symbols  and model grid lines are the same as in Fig.~\ref{fig:M32_4m}.  In
addition, we plot as large open circles the spectral indices obtained from
adding varying amounts of a simulated featureless scattered light spectrum to
the observed M32 nuclear spectrum.  Bluer colored circles denote less scattered
light component, while red indicates greater scattered light.  This simple
scattered light model appears to track the observed gradient in M32 indices.
}
\label{fig:scatter}
\end{figure}

\begin{figure}
\plotone{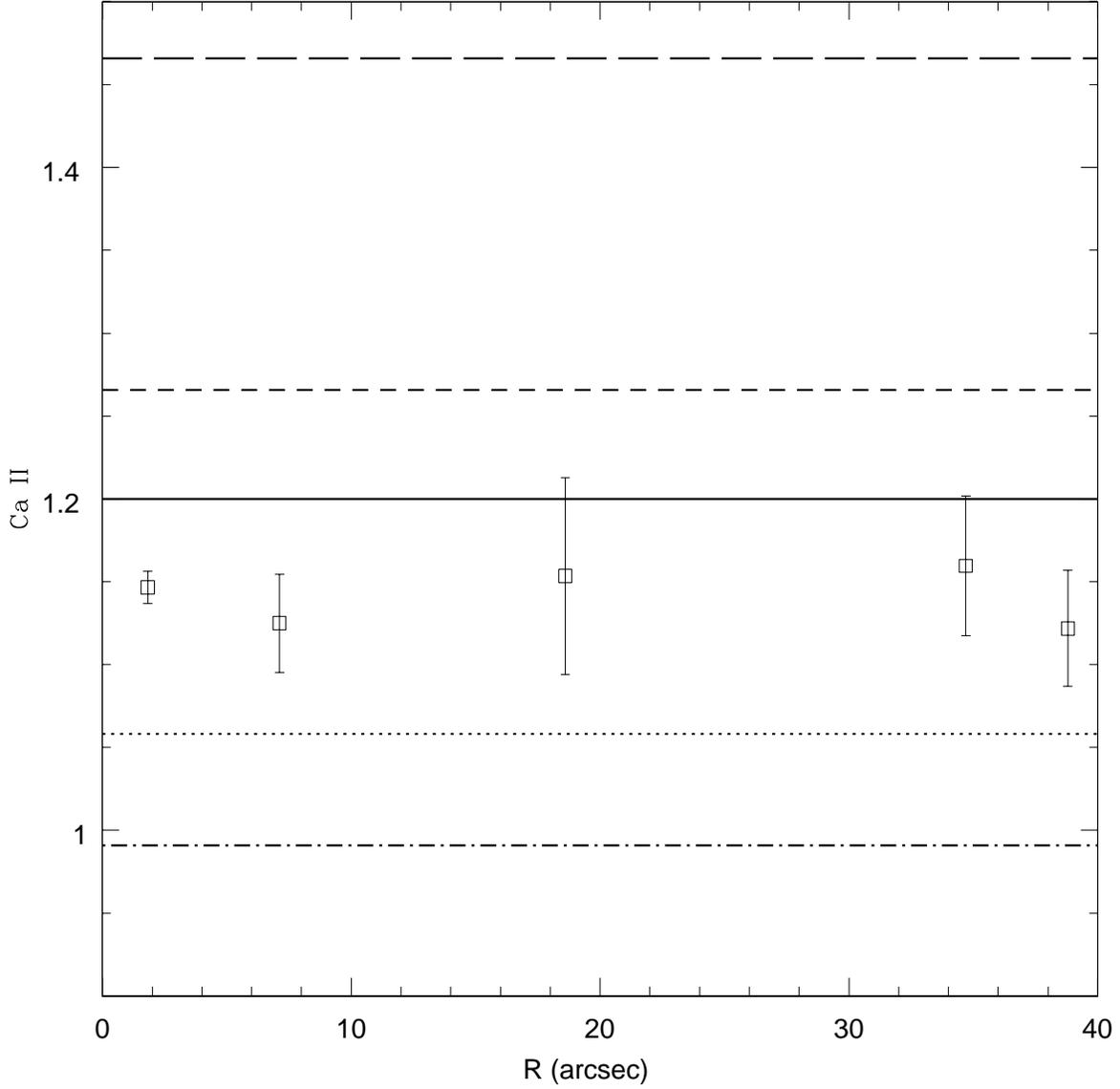}
\caption{The Ca II index is plotted versus radial distance in M32 for the FAST
data.  The data are plotted as open squares with $\pm$1$\sigma$ error bars.
The solid line at at Ca II value of 1.2 shows the expected value for an
entirely cool star population.  The dotted and dot-dash lines represent the
Ca II values found when the spectrum of an A4V star is added to that of M32 at
the 5\% and 10\% levels respectively, with normalization at 4000 \AA.  The
short and long dashed lines represent the values found when the A4V spectrum
is subtracted off at 5\% and 10\% levels.}
\label{fig:caII}
\end{figure}

\begin{figure}
\plotone{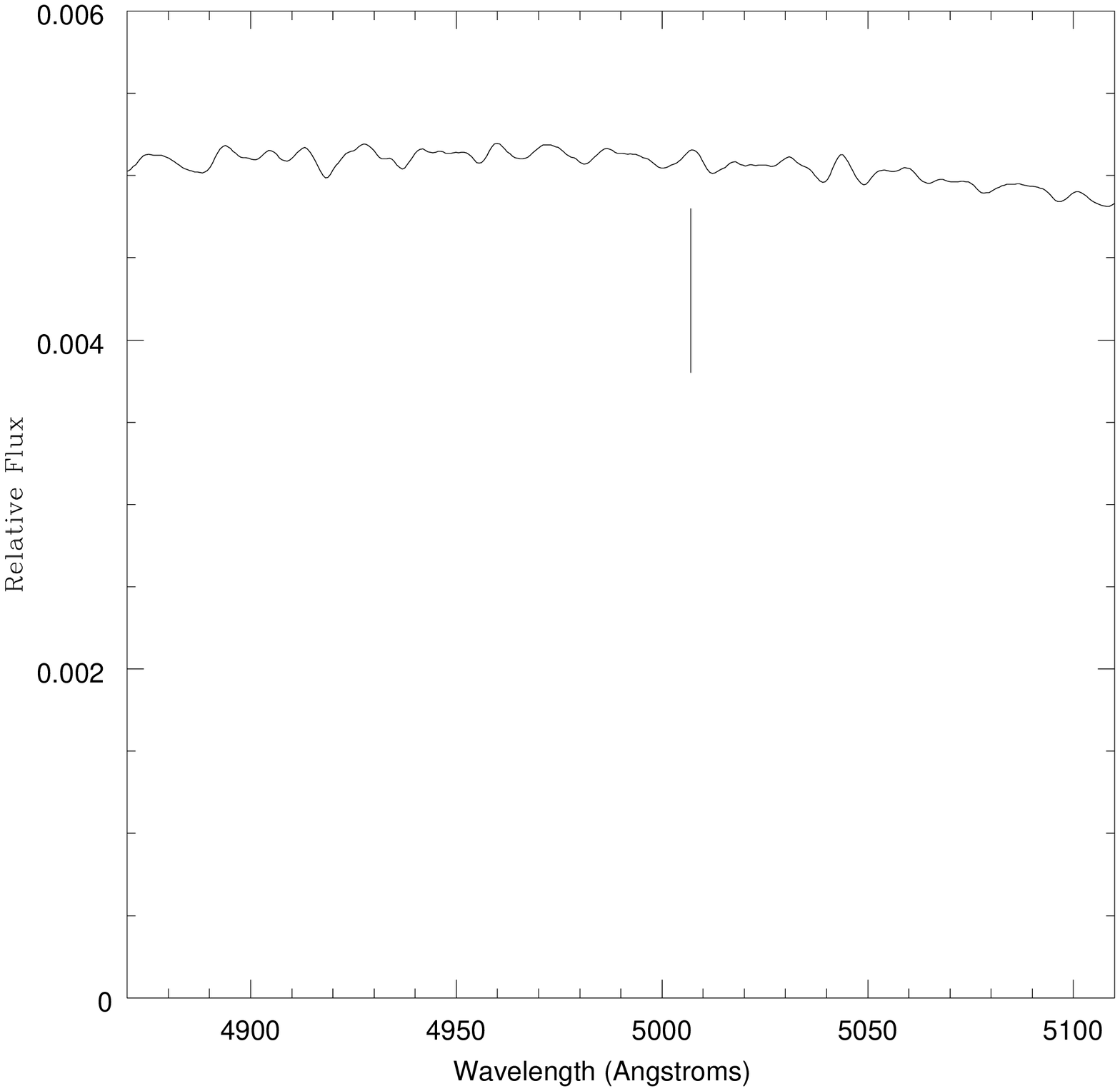}
\caption{Ratio spectrum of the nucleus of M32 divided by the integrated spectrum
of M67 is plotted in the region of the [OIII]$\lambda$5007 emission line.
The expected location of [OIII]$\lambda$5007 emission is marked.}
\label{fig:emis5007}
\end{figure}

\begin{figure}
\plotone{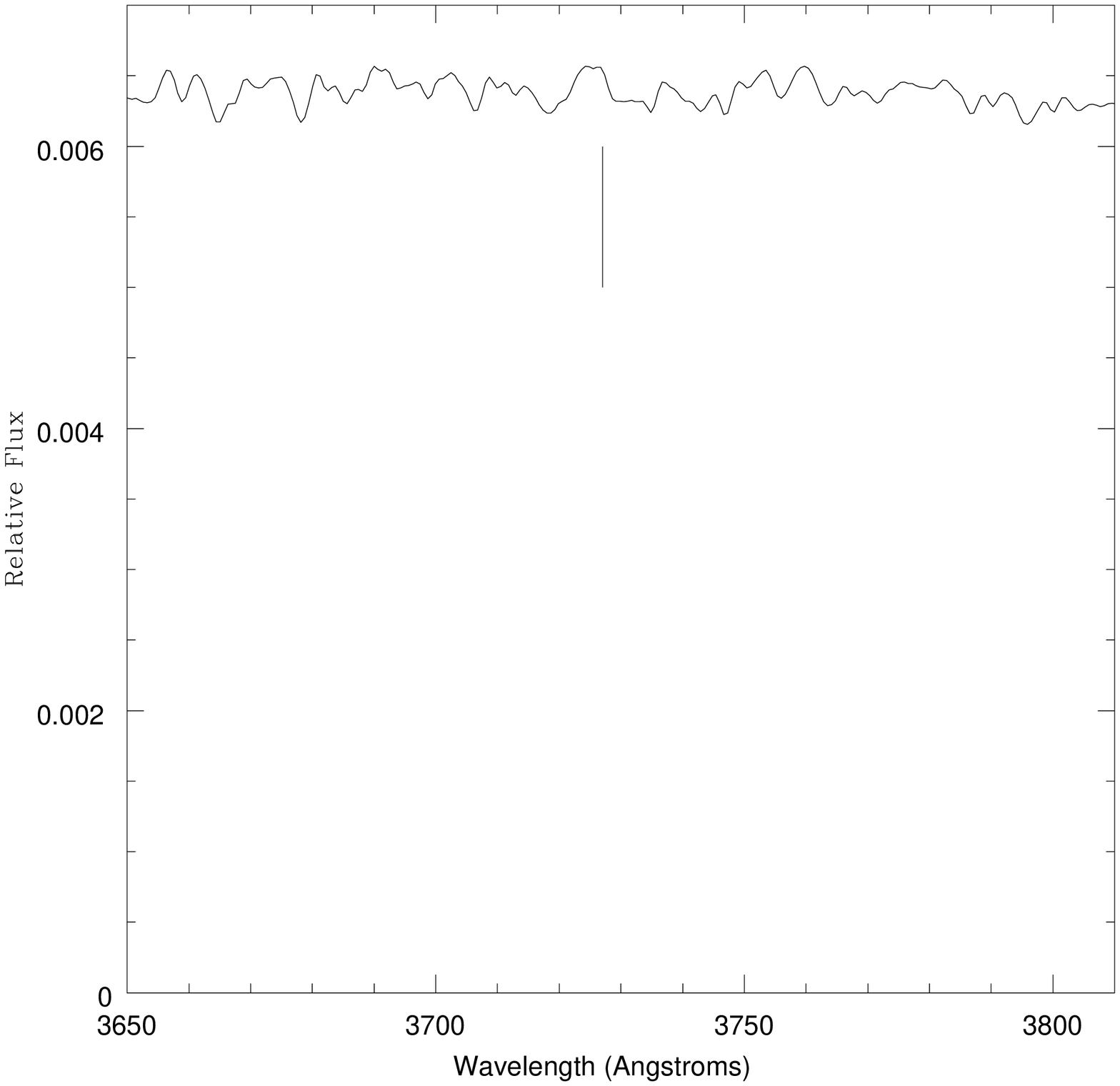}
\caption{Ratio spectrum of the nucleus of M32 divided by the integrated spectrum
of M67 is plotted in the region of the [OII]$\lambda$3727 emission line.
The expected location of [OII]$\lambda$3727 emission is marked.}
\label{fig:emis3727}
\end{figure}

As was mentioned in the Introduction, no gradient in broad passband colors has
been found for the central regions of M32 in either the optical or IR
(Peletier 1993), which appears to contradict the distinct age/metallicity
gradient found in our spectra.  However, the trend is towards older age and
lower metal-abundance as one proceeds radially outward.  An increase in age
produces a reddening of colors while the decrease in metallicity has the
opposite effect.  Thus we have the interesting situation in which the
competing effects of age and metallicity lead to no variation in optical/IR
colors.  To investigate the plausibility of this assertion, we have assessed
the B$-$V and U$-$B colors predicted from our stellar population models.
The model B$-$V color for a 4 Gyr solar abundance population is 0.89, as opposed
to 0.86 for a 7 Gyr, [Fe/H]=$-$0.3 population.  Thus the radial change in
color is very small and difficult to detect.  For U$-$B the model color shifts
from 0.44 to 0.37 from the younger to the older population, which is again a
small shift, but perhaps should be detectable.  We note, however, that our
models are based on the clear oversimplification of a single population  at
each radius that is entirely at the luminosity-weighted mean value, while in
reality there must  be a spread in both age and metallicity.  In addition, our
models are basically limited to solar abundance ratios, while our data for M32
indicates departures from that pattern.  In fact, while the model B$-$V color
for the center of M32 is consistent with the observed color, to within the
uncertainties of Galactic reddening in that direction, the model U$-$B colors
are systematically bluer than the observed colors reported in Peletier (1993)
by $\sim$0.1$-$0.15.
We also note that some spectral features, e.g.,
Mg~$b$, also show no significant line strength gradient, again because an
older population tends to have stronger Mg~$b$, while a more metal-poor
population tends to have weaker Mg~$b$, as can be seen in
Fig.~\ref{fig:M32Sub3}.  Hence the most striking line strength
gradients that we observe are in the Balmer lines.

It is important to bear in mind that the above inferences about radial
population trends are for the
light-weighted mean population.  An alternative scenario to the possibility
of an overall shift of the bulk of the population to a younger age, as one
proceeds to smaller radii, is to invoke a shifting balance between two
populations.  Specifically, one can speculate that there was an initial star
formation episode $\sim$11 Gyr ago, as suggested by the presence of blue
horizontal branch stars in color-magnitude diagrams derived from HST images
(Ikuta 2001; Alonso-Garcia \etal 2003), as well as a more recent, higher 
metallicity, star formation episode perhaps 2$-$3 Gyr ago.  The observed population 
trend can then be seen as a higher concentration of the younger population in 
the center of M32.  Such a population structure would in fact be a natural 
result of the scenario proposed by Bekki \etal (2001), in which centralized
star formation in M32 is triggered by the tidal field of M31.
To discriminate between a mean trend and a changing balance between two
populations is beyond the scope of this paper.  It should be kept in mind
that the more complex modelling involved with a two-population scenario could
as well impact the details of conclusions about the small gradient in
non-solar abundance ratios (viz., in [Mg/Fe]) found earlier.
Note that in any case
the younger population cannot be any younger than $\sim$2 Gyr, or it
would produce a strong signature in the Ca II index discussed in 
\S\ref{hotstars}.

Finally, we return to the issue of scattered light in M32.  In the Introduction
we mentioned the conflicting results of several previous long-slit spectroscopic
studies of M32.  Most, if not all, of these studies involve spectra taken with
the slit passing through the nucleus of M32, hence they are particularly
vulnerable to effects of scattered light.  Perhaps the major discrepancies
between studies are due to varying degrees of scattered light.
In addition, while M32 represents an especially pathological case of
a bright nucleus as a source of scattered light, it is not inconceivable that
studies of radial population gradients in other early-type galaxies can have
been compromised by scattered light as well.

\acknowledgements

We wish to thank S. M. Faber for the crucial suggestion of scattered
light as a source of spurious age/metallicity gradients in M32.  Thanks are
also due to Dr. R. Peletier and Drs. C. Ikuta for their efforts in securing
the Subaru/FOCAS observations.
This research has been partially supported by NSF grant AST-9900720 to the
University of North Carolina.  It has also been supported in part by a 
Grant-in-Aid for Scientific Research by the Japanese Ministry of Education,
Culture, Sports, Science and Technology (No. 13640230).

\newpage

\begin{deluxetable}{lllccccll}
\tabletypesize{\footnotesize}
\tablewidth{0pt}
\tablenum{1}
\tablecolumns{9}

\tablecaption{Journal of Observations\label{tab:journal}}
\tablehead{
\colhead{UT Date} &
\colhead{Telescope} &
\colhead{Slit} &
\colhead{PA} &
\colhead{Exposure } &
\colhead{Slit} &
\colhead{Extraction} &
\colhead{Wavelength} &
\colhead{Resolution} \\
\colhead{} &
\colhead{} &
\colhead{Position} &
\colhead{} &
\colhead{Time} &
\colhead{Width} &
\colhead{Area} &
\colhead{Coverage} &
\colhead{FWHM}
}

\startdata

12$/$28$/$2000 & Tillinghast + FAST & M32 Nucleus & 90 & 8 x 240s & 3\arcs & $\pm$5.7\arcs & 3600 - 5500 \AA & 3.1 \AA \\
12$/$28$/$2002 & Tillinghast + FAST & M32 Offset 5\arcs & -20 & 900s & 3\arcs & $\pm$9.1\arcs & \nodata & \nodata \\
12$/$28$/$2002 & Tillinghast + FAST & M32 Offset 15\arcs & -20 & 2 x 900s & 3\arcs & $\pm$22.8\arcs & \nodata & \nodata \\
12$/$28$/$2002 & Tillinghast + FAST & Sky Offset 10\arcm S & -20 & 900s & 3\arcs & \nodata & \nodata & \nodata \\
12$/$28$/$2002 & Tillinghast + FAST & M32 Offset 30\arcs  & -20 & 7 x 900s & 3\arcs & $\pm$34.2\arcs & \nodata & \nodata \\
\nodata &\nodata & \nodata& \nodata& \nodata& \nodata& $\pm$54.7 & \nodata & \nodata \\
12$/$28$/$2002 & Tillinghast + FAST & Sky Offset 10\arcm S & -20 & 4 x 900s & 3\arcs & \nodata & \nodata & \nodata \\
 & & & & & & & & \\
01$/$28$/$2003 & Subaru + FOCAS & M32 Nucleus & 80 & 2 x 300s & 0.6\arcs & $\pm$3\arcs & 3960 - 5560 \AA & 3.1 \AA \\
01$/$28$/$2003 & Subaru + FOCAS & M32 Offset by 10\arcs & 80 & 3 x 900s & 0.6\arcs & \nodata\tablenotemark{a} & \nodata & \nodata \\
01$/$28$/$2003 & Subaru + FOCAS & M32 Sky Offset 6\arcm E & 80 & 900s & 0.6\arcs & \nodata & \nodata & \nodata \\
 & & & & & & & & \\
06$/$18$/$1996 & Mayall + R-C Spectro & M32 Nucleus & 102 & 1800s & 2\arcs & \nodata\tablenotemark{b} & 3300 - 4770 \AA & 3.1 \AA \\
06$/$18$/$1996 & Mayall + R-C Spectro & Sky Offset by  & 102 & 1800s & 2\arcs & \nodata & \nodata & \nodata \\
06$/$19$/$1996 & Mayall + R-C Spectro & M32 Nucleus & 100 & 1800s & 2\arcs & \nodata\tablenotemark{b} & \nodata & \nodata \\
06$/$19$/$1996 & Mayall + R-C Spectro & Sky Offset by  & 100 & 1800s & 2\arcs & \nodata & \nodata & \nodata \\
06$/$20$/$1996 & Mayall + R-C Spectro & M32 Nucleus & 155 & 2 x 1800s & 2\arcs & \nodata\tablenotemark{b} & \nodata & \nodata \\
06$/$20$/$1996 & Mayall + R-C Spectro & Sky Offset by  & 155 & 1800s & 2\arcs & \nodata & \nodata & \nodata \\

\enddata
\tablenotetext{a} {Extractions were made at 10$-$14\arcsec, 14.1$-$31.6\arcsec,
24.2$-$39.3\arcs}
\tablenotetext{b} {Extractions were made at 0$-$1.4\arcsec, 2.8$-$5.5\arcsec,
6.2$-$11.0\arcsec, 11.7$-$26.9\arcsec, 27.6$-$41.4\arcs}

\end{deluxetable}

\begin{deluxetable}{rrrrrrrrrrrrrr}
\tabletypesize{\footnotesize}
\tablewidth{0pt}
\tablenum{2}
\tablecolumns{13}

\tablecaption{Spectral Indices and Errors for Subaru Data\tablenotemark{a}
\label{tab:Subaru}}
\tablehead{
\colhead{Radius} &
\colhead{Radius} &
\colhead{H$\gamma_{\sigma<130}$} &
\colhead{H$\gamma_{125}$} &
\colhead{H$\delta_A$} &
\colhead{H$\gamma_A$} &
\colhead{H$\delta_F$} &
\colhead{H$\gamma_F$} &
\colhead{CN1} &
\colhead{CN2} &
\colhead{Ca4226} &
\colhead{G4300} &
\colhead{Fe4383} \\
\colhead{''} &
\colhead{R$_e$} &
\colhead{$\pm1\sigma$} &
\colhead{$\pm1\sigma$} &
\colhead{$\pm1\sigma$} &
\colhead{$\pm1\sigma$} &
\colhead{$\pm1\sigma$} &
\colhead{$\pm1\sigma$} &
\colhead{$\pm1\sigma$} &
\colhead{$\pm1\sigma$} &
\colhead{$\pm1\sigma$} &
\colhead{$\pm1\sigma$} &
\colhead{$\pm1\sigma$} \\
\colhead{} &
\colhead{} &
\colhead{H$\beta$} &
\colhead{Fe5015} &
\colhead{Mg$_1$} &
\colhead{Mg$_2$} &
\colhead{Mg~$b$} &
\colhead{Fe5270} &
\colhead{Fe5335} &
\colhead{Fe5406} &
\colhead{[MgFe]} &
\colhead{Fe3} &
\colhead{Ca II} \\
\colhead{} &
\colhead{} &
\colhead{$\pm1\sigma$} &
\colhead{$\pm1\sigma$} &
\colhead{$\pm1\sigma$} &
\colhead{$\pm1\sigma$} &
\colhead{$\pm1\sigma$} &
\colhead{$\pm1\sigma$} &
\colhead{$\pm1\sigma$} &
\colhead{$\pm1\sigma$} &
\colhead{$\pm1\sigma$} &
\colhead{$\pm1\sigma$} &
\colhead{$\pm1\sigma$} 
}

\startdata

0.7    & 0.02    &   1.261 &   1.055 &  -1.178 &  -4.342 &   0.996 &  -0.629 &   0.023 &   0.063 &   1.242 &   5.013 &   5.364 \\ 
       &        &   0.022 &   0.018 &   0.032 &   0.027 &   0.021 &   0.016 &   0.001 &   0.001 &   0.014 &   0.024 &   0.032 \\ 
       &        &   2.367 &   5.859 &   0.046 &   0.159 &   3.032 &   3.271 &   3.131 &   2.021 &   3.115 &   3.922 & \nodata \\ 
       &        &   0.014 &   0.031 &   0.000 &   0.000 &   0.016 &   0.017 &   0.020 &   0.015 &   0.007 &   0.014 &  \nodata \\
10.3   & 0.29   &   1.224 &   1.045 &  -1.324 &  -4.421 &   0.846 &  -0.757 &   0.019 &   0.058 &   1.348 &   5.224 &   5.168 \\ 
       &        &   0.033 &   0.028 &   0.051 &   0.041 &   0.033 &   0.026 &   0.001 &   0.001 &   0.022 &   0.037 &   0.050 \\ 
       &        &   2.231 &   5.696 &   0.046 &   0.161 &   3.131 &   3.241 &   3.070 &   1.953 &   3.143 &   3.826 & \nodata \\ 
       &        &   0.022 &   0.047 &   0.001 &   0.001 &   0.024 &   0.026 &   0.030 &   0.023 &   0.011 &   0.021 &  \nodata \\
20.8   & 0.59   &   1.182 &   1.026 &  -1.126 &  -4.413 &   0.766 &  -0.822 &   0.008 &   0.045 &   1.256 &   5.188 &   5.107 \\ 
       &        &   0.038 &   0.031 &   0.057 &   0.046 &   0.038 &   0.029 &   0.001 &   0.002 &   0.025 &   0.041 &   0.056 \\ 
       &        &   2.099 &   5.677 &   0.046 &   0.162 &   3.207 &   3.192 &   3.044 &   1.957 &   3.162 &   3.781 &  \nodata \\
       &        &   0.024 &   0.053 &   0.001 &   0.001 &   0.026 &   0.029 &   0.034 &   0.026 &   0.012 &   0.024 &  \nodata \\
30.3   & 0.86   &   1.145 &   0.947 &  -1.186 &  -4.265 &   0.488 &  -0.869 &  -0.004 &   0.033 &   1.177 &   4.876 &   4.733 \\ 
       &        &   0.057 &   0.047 &   0.086 &   0.070 &   0.058 &   0.044 &   0.002 &   0.003 &   0.038 &   0.063 &   0.085 \\ 
       &        &   2.001 &   5.486 &   0.043 &   0.158 &   3.123 &   3.042 &   2.931 &   1.933 &   3.054 &   3.569 &  \nodata \\
       &        &   0.037 &   0.080 &   0.001 &   0.001 &   0.040 &   0.044 &   0.051 &   0.039 &   0.019 &   0.036 &  \nodata \\
\enddata
\tablenotetext{a}{All spectral indices have been determined after first smoothing
the spectra to a final effective velocity dispersion of 130 \kms, when 
compared to a template with intrinsic spectral resolution of 1.8 \AA \ FWHM.
This corresponds to a final spectral resolution of 4.75 \AA \ FWHM at 4300 \AA.
Note that our spectral resolution is not that of the Lick/IDS system
resolution.  As well, the indices are measured from the flux calibrated spectra,
i.e., not the Lick/IDS instrumental profile.}

\end{deluxetable}

\begin{deluxetable}{rrrrrrrrrrrrrr}
\tabletypesize{\footnotesize}
\tablewidth{0pt}
\tablenum{3}
\tablecolumns{14}

\tablecaption{Spectral Indices and Errors for Tillinghast Data\tablenotemark{a}
\label{tab:FAST}}
\tablehead{
\colhead{Radius} &
\colhead{Radius} &
\colhead{H$\gamma_{\sigma<130}$} &
\colhead{H$\gamma_{125}$} &
\colhead{H$\delta_A$} &
\colhead{H$\gamma_A$} &
\colhead{H$\delta_F$} &
\colhead{H$\gamma_F$} &
\colhead{CN1} &
\colhead{CN2} &
\colhead{Ca4226} &
\colhead{G4300} & 
\colhead{Fe4383} \\
\colhead{''} &
\colhead{R$_e$} &
\colhead{$\pm1\sigma$} &
\colhead{$\pm1\sigma$} &
\colhead{$\pm1\sigma$} &
\colhead{$\pm1\sigma$} &
\colhead{$\pm1\sigma$} &
\colhead{$\pm1\sigma$} &
\colhead{$\pm1\sigma$} &
\colhead{$\pm1\sigma$} &
\colhead{$\pm1\sigma$} &
\colhead{$\pm1\sigma$} &
\colhead{$\pm1\sigma$} \\
\colhead{} &
\colhead{} &
\colhead{H$\beta$} &
\colhead{Fe5015} &
\colhead{Mg$_1$} &
\colhead{Mg$_2$} &
\colhead{Mg~$b$} &
\colhead{Fe5270} &
\colhead{Fe5335} &
\colhead{Fe5406} &
\colhead{[MgFe]} &
\colhead{Fe3} &
\colhead{Ca II} \\
\colhead{} &
\colhead{} &
\colhead{$\pm1\sigma$} &
\colhead{$\pm1\sigma$} &
\colhead{$\pm1\sigma$} &
\colhead{$\pm1\sigma$} &
\colhead{$\pm1\sigma$} &
\colhead{$\pm1\sigma$} &
\colhead{$\pm1\sigma$} &
\colhead{$\pm1\sigma$} &
\colhead{$\pm1\sigma$} &
\colhead{$\pm1\sigma$} &
\colhead{$\pm1\sigma$} &
\colhead{}
}

\startdata

2.1 & 0.06 &   1.266 &   1.110 &  -1.097 &  -4.472 &   0.984
&  -0.712 &   0.031 &   0.073 &   1.395 &   5.438  &   5.458 \\
 &  &   0.014 &   0.011 &   0.068 &   0.068 &   0.073
&   0.033 &   0.002 &   0.003 &   0.051 &   0.051 &   0.110 \\
 & &   2.330 &   5.815 &   0.092 &   0.211 &   3.066 &   3.208 &   3.0
03 &   2.002 &   3.086 &   3.890 & 1.147 \\
 & &   0.063 &   0.101 &   0.001 &   0.002 &   0.034 &   0.051 &   0.0
56 &   0.034 &   0.018 &   0.045 & 0.010 \\
7.1 & 0.24 &   1.214 &   1.060 &  -0.998 &  -4.283 &   0.923 &  -0.575 &   0.039 &   0.082 &   1.424 &   5.308 &   5.529 \\
       &        &   0.071 &   0.058 &   0.101 &   0.085 &   0.067 &   0.054 &   0.003 &   0.003 &   0.045 &   0.076 &   0.105 \\
       &        &   2.274 &   5.617 &   0.081 &   0.194 &   3.083 &   3.300 &   2.959 &   1.896 &   3.107 &   3.930 & 1.125 \\
       &        &   0.043 &   0.091 &   0.001 &   0.001 &   0.044 &   0.048 &   0.054 &   0.040 &   0.019 &   0.042 & 0.030 \\
18.6 & 0.64 &   1.103 &   0.955 &  -0.876 &  -3.969 &   0.964 &  -0.640 &   0.038 &   0.085 &   1.286 &   5.466 &   4.811 \\
       &        &   0.060 &   0.050 &   0.086 &   0.073 &   0.057 &   0.046 &   0.002 &   0.003 &   0.038 &   0.065 &   0.089 \\
       &        &   2.234 &   5.449 &   0.081 &   0.199 &   3.213 &   3.057 &   2.758 &   1.900 &   3.057 &   3.542 & 1.154 \\
       &        &   0.038 &   0.080 &   0.001 &   0.001 &   0.038 &   0.041 &   0.047 &   0.035 &   0.018 &   0.036 & 0.059 \\
34.7 & 1.19 &   1.011 &   0.889 &  -1.151 &  -4.152 &   0.795 &  -0.731 &   0.029 &   0.071 &   1.422 &   5.429 &   5.088 \\
       &        &   0.043 &   0.036 &   0.059 &   0.052 &   0.040 &   0.033 &   0.002 &   0.002 &   0.027 &   0.047 &   0.065 \\
       &        &   2.150 &   5.082 &   0.078 &   0.195 &   3.112 &   3.059 &   2.818 &   1.349 &   3.024 &   3.655 & 1.160 \\
       &        &   0.027 &   0.058 &   0.001 &   0.001 &   0.028 &   0.029 &   0.031 &   0.023 &   0.012 &   0.026 & 0.042 \\
38.8 & 1.33 &   1.136 &   1.021 &  -0.833 &  -4.143 &   0.718 &  -0.732 &   0.020 &   0.059 &   1.337 &   5.266 &   4.872 \\
       &        &   0.048 &   0.040 &   0.067 &   0.059 &   0.045 &   0.037 &   0.002 &   0.002 &   0.031 &   0.052 &   0.072 \\
       &        &   2.099 &   4.955 &   0.075 &   0.191 &   3.067 &   2.961 &   2.585 &   1.337 &   2.916 &   3.473 & 1.122 \\
       &        &   0.030 &   0.065 &   0.001 &   0.001 &   0.031 &   0.033 &   0.037 &   0.026 &   0.014 &   0.029 & 0.035 \\

\enddata
\tablenotetext{a}{All spectral indices have been determined after first smoothin
g
the spectra to a final effective velocity dispersion of 130 \kms, when
compared to a template with intrinsic spectral resolution of 1.8 \AA \ FWHM.
This corresponds to a final spectral resolution of 4.75 \AA \ FWHM at 4300 \AA.
Note that our spectral resolution is not that of the Lick/IDS system
resolution.  As well, the indices are measured from the flux calibrated spectra,
i.e., not the Lick/IDS instrumental profile.}

\end{deluxetable}


\begin{references}

\reference{amb99} Abadi, M. G., Moore, B., \& Bower, R. G. 1999, \mnras, 308,
947
\reference{al96} Alonso, A., Arribas, S., Martinez-Roger, C, 1996, \aap, 313, 873
\reference{al99} Alonso, A., Arribas, S., Martinez-Roger, C, 1999, \aap, 140, 261
\reference{ag03} Alonso-Garcia, J., Mateo, M., \& Worthey, G. 2003, \aj, 127, 868
\reference{bp94} Balcells, M., \& Peletier, R. F. 1994, \aj, 107, 135
\reference{bek01} Bekki, K., Couch, W. J., Drinkwater, M. J., \& Gregg, M. D.
2001, \apj, 557, L39
\reference{be94} Bertelli, G., Bressan, A., Chiosi, C., Fagotto, F., \& Nasi, E.
1994, \aaps, 106, 275
\reference{br98} Brown, T. M., Ferguson, H. C., Stanford, S. A., \& Deharveng,
J.-M. 1998, \apj, 504, 113
\reference{br00} Brown, T. M., Bowers, C. W., Kimble, R. A., \& Sweigart, A. V.
2000, \apj, 532, 308
\reference{bc03} Bruzual, G., \& Charlot, S. 2003, \mnras, 344, 1000
\reference{bu85} Burstein, D. 1985, \pasp, 97, 89
\reference{bu84} Burstein, D., Faber, S. M., Gaskell, C. M., \& Krumm, N. 1984,
\apj, 287, 586
\reference{crc} Caldwell, N., Rose, J. A., \& Concannon, K. D. 2003, \aj, 125, 2891
\reference{ca98} Cardiel, N., Gorgas, J., Cenarro, J., \& Gonz\'alez, J. J. 1998,
\aaps, 127, 597
\reference{ci89} Ciardullo, R., Jacoby, G. H., Ford, H. C., \& Neill, J. D.
1989, \apj, 339, 53
\reference{ch02} Choi, P. I., Guhathakurta, P., Johnston, K. V. 2002, \aj, 124, 310
\reference{jc79} Cohen, J. 1979, \apj, 228, 405
\reference{td91} Davidge, T. J. 1991, \aj, 101, 884
\reference{td00} Davidge, T. J. 2000, \pasp, 112, 1177
\reference{tdb00} Davidge, T. J., Rigaut, F., Chun, M., Brandner, W., Potter,
D., Northcutt, M., \& Graves, J. E. 2000, \apj, 545, L89
\reference{td90} Davidge, T. J., de Robertis, M. M., \& Yee, H. K. C. 1990,
\aj, 100, 1143
\reference{dj92} Davidge, T. J., \& Jones, J. H. 1992, \aj, 104, 1365
\reference{dn92} Davidge, T. J., \& Nieto, J.-L. 1992, \apj, 391, L13
\reference{del00} del Burgo, C., Peletier, R. F., Vazdekis, A., Arribas, S., \&
Mediavilla, E. 2001, \mnras, 321, 227
\reference{dep00} de Propris, R. 2000, \mnras, 316, L9
\reference{dev48} de Vaucouleurs, G. 1948, Ann. d'Ap., 11, 247
\reference{dr88} Dressler, A., \& Richstone, D. O. 1988, \apj, 324, 701
\reference{es92} Elston, R., \& Silva, D. R. 1992, \aj, 104, 1360
\reference{fa98} Fabricant, D., Cheimets, P., Caldwell, N., \& Geary, J. 1998,
\pasp, 110, 79
\reference{ffi95} Fisher, D., Franx, M., \& Illingworth, G. 1995, \apj, 447,
L139
\reference{wf89} Freedman, W. L. 1989, \aj, 98, 1285
\reference{wf92} Freedman, W. L. 1992, \aj, 104, 1349
\reference{gdp03} Gil de Paz, A., Madore, B. F., Rich. M., Seibert, M., \&
GALEX Science Team 2003, \baas, 203, 9606
\reference{gi00} Girardi, L., Bressan, A., Bertelli, G., \& Chiosi, C. 2000, 
\aaps, 141, 371
\reference{jjg93} Gonz\'alez, J. J. PhD Thesis, University of California, Santa Cruz
\reference{go93} Gorgas, J., Faber, S. M., Burstein, D., Gonz\'alez, J. J., 
Courteau, S., \& Prosser, C. 1993, \apjs, 86, 153
\reference{gr04} Gregg, M. D., Ferguson, H. C., Minniti, D., Tanvir, N., \&
Catchpole, R. 2004, \aj, 127, 1441
\reference{gr96} Grillmair, C. J., Lauer, T. R., Worthey, G., Faber, S. M.,
Freedman, W. L., Madore, B. F., Ajhar, E. A., Baum, W. A., Holtzman, J. A.,
Lynds, C. R., O'Neil, E. J., Jr., \& Stetson, P. B. 1996, \aj, 112, 1975
\reference{har94} Hardy, E., Couture, J., Couture, C., Joncas, G. 1994, \aj,
107, 195
\reference{jo99} Jones, L. A. 1999, PhD Thesis, University of North Carolina
\reference{jw95} Jones, L. A., \& Worthey, G. 1995, \apj, 446, L31
\reference{ka00} Kashikawa, N. \etal 2000, SPIE, 4008, 104
\reference{ke87} Kent, S. M. 1987, \aj, 94, 306
\reference{koa99} Kobayashi, C., \& Arimoto, N. 1999, \apj, 527, 573
\reference{tl98} Lauer, T. R., Faber, S. M., Ajhar, E. A., Grillmair, 
C. J., \& Scowen, P. A. 1998, \aj, 116, 2263
\reference{leo00} Leonardi, A. J., \& Rose. J. A. 2003, \aj, 126, 1811
\reference{mt00} Maraston, C., \& Thomas, D. 2000, \apj, 541, 126
\reference{ohl98} Ohl, R. G., O'Connell, R. W., Bohlin, R. C., Collins, N. R.,
Dorman, B., Fanelli, M. N., Neff, S. G., Roberts, M. S., Smith, A. M., \&
Stecher, T. P. 1998, \apj, 505, L11
\reference{rp93} Peletier, R. F. 1993, \aap, 271, 51
\reference{ro84} Rose, J. A. 1984, \aj, 89, 1238
\reference{ro85} Rose, J. A. 1985, \aj, 90, 1927
\reference{ro94} Rose, J. A. 1994, \aj, 107, 206
\reference{rd99} Rose, J. A., \& Deng, S. 1999, \aj, 117, 2213
\reference{sch04} Schiavon, R. P. 2004, in preparation
\reference{sch00} Schiavon, R. P., Barbuy, B., \& Bruzual, A. G. 2000, \apj,
532, 453
\reference{scr04} Schiavon, R. P., Caldwell, N., \& Rose, J. A. 2004, \aj, 127, 1513
\reference{ss72} Searle, L., \& Sargent, W. L. W. 1972, \apj, 173, 611
\reference{tmb03} Thomas, D., Maraston, C., \& Bender, R. 2003, \mnras, 339, 897
\reference{to84} Tonry, J. L. 1984, \apj, 283, L27
\reference{tr00} Trager, S. C., Faber, S. M., Worthey, G., \& Gonz\'alez, J. J.
2000, 120, 165
\reference{tb95} Tripicco, M. J., \& Bell, R. A. 1995, \aj, 110, 3035
\reference{va04} Valdes, F., Gupta, R., Rose, J. A., Singh, H. P., \& Bell, D.
J. 2004, \apjs, 152, 251
\reference{vdm94} van der Marel, R. P., Evans, N. W., Rix, H.-W., White, S. D. 
M., \& de Zeeuw, T. 1994, \mnras, 271, 99
\reference{vdm98} van der Marel, R. P., Cretton, N., de Zeeuw, P. T., \& Rix, 
H.-W. 1998, \apj, 493, 613
\reference{va96} Vazdekis. A., Casuso, E., Peletier, R. F., \& Beckman, J. E.
1996, \apj, 458, 533
\reference{vaz99} Vazdekis, A. 1999, \apj, 513, 224
\reference{va99} Vazdekis, A., \& Arimoto, N. 1999, \apj, 525, 144
\reference{va01a} Vazdekis, A., Kuntschner, H., Davies, R. L., Arimoto, N.,
Nakamura, O., \& Peletier, R. 2001a, \apj, 551, L127
\reference{va01b} Vazdekis, A., Salaris, M., Arimoto, N., \& Rose, J. A.
2001b, \apj, 549, 274
\reference{va03} Vazdekis, A., Cenarro, A. J., Gorgas, J., Cardiel, N.,
Peletier, R. F. 2003, \mnras, 340, 1317
\reference{ve02} Verolme, E. K., Cappellari, M., Copin, Y., van der Marel, R. P., Bacon, R.,
Bureau, M., Davies, R. L., Miller, B. M., de Zeeuw, P. T. 2002, \mnras, 335, 517
\reference{we02} Westera, P., Lejeune, T., Buser, R., Cuisinier, F., \& 
Bruzual, G. 2002, \aap, 381, 524
\reference{wor94} Worthey, G. 1994, \apjs, 95, 107
\reference{wo03} Worthey, G. 2004, \aj, (submitted)
\reference{wo94} Worthey, G., Faber, S. M., Gonz\'alez, J. J., \& Burstein, D. 
1994, \apjs, 94, 687
\reference{wo97} Worthey, G., \& Ottaviani, D. L. 1997, \apjs, 111, 377
White, S. D. M., \& de Zeeuw, T., 1994, \mnras, 268, 521


\end{references}
\end{document}